\documentclass[conference]{IEEEtran}

\usepackage{amsmath}
\usepackage{amsfonts}
\usepackage{amsthm}
\usepackage{xcolor}
\usepackage{hyperref}
\usepackage[capitalize,noabbrev]{cleveref}
\usepackage[breakable]{tcolorbox}
\usepackage[detect-all=true]{siunitx}
\usepackage{subcaption}
\usepackage[mode=buildnew]{standalone}
\usepackage{doi}
\usepackage{balance}
\usepackage{bm}
\usepackage{orcidlink}

\tcbuselibrary{breakable}
\newcommand{\greyborder}[1]{%
  \begin{tcolorbox}[breakable, colback=white, colframe=darkgray, left=1pt,right=10pt,top=0pt,bottom=0pt]%
    #1%
  \end{tcolorbox}%
}

\newcommand{\geopkiorig}{GeoPKI}

\newcommand{\fgpki}{GECKO}
\newcommand{\fgpkifull}{Geo-Enabled Cryptographic Key Oracle}
\newcommand{\fgpkischeme}{gecko}

\newcommand{\fpkiorig}{F-PKI}
\newcommand{\geocert}{GeoCert}
\newcommand{\geocerts}{GeoCerts}

\newcommand{\webpkideployment}{web PKI extension deployment}
\newcommand{\standalonedeployment}{standalone deployment}

\newcommand{\myparagraph}[1]{\noindent\textbf{#1.}\quad}

\DeclareMathAlphabet{\mymathbb}{U}{BOONDOX-ds}{m}{n} % mathbb zero and one
\newcommand{\mathbbzero}{\mymathbb{0}}
\newcommand{\mathbbone}{\mymathbb{1}}

\usepackage{scalerel}
\DeclareMathOperator*{\concat}{\scalerel*{||}{\sum}}

\usepackage[backend=biber,giveninits=true,maxbibnames=6,sorting=none,doi=true,isbn=true]{biblatex}
\DeclareFieldFormat[article]{volume}{\bibstring{jourvol}\addnbspace #1}
\DeclareFieldFormat[article]{number}{\bibstring{number}\addnbspace #1}
\renewbibmacro*{volume+number+eid}{%
  \printfield{volume}%
  \newunit%
  \printfield{number}%
  \newunit%
  \printfield{eid}}

\title{\fgpki{}: Securing Digital Assets Through(out) the Physical World (Extended Technical Report)}

\author{\IEEEauthorblockN{Cyrill Krähenbühl \orcidlink{0000-0001-9988-1271}}
\IEEEauthorblockA{\textit{Princeton University}}
\and
\IEEEauthorblockN{Nico Hauser}
\IEEEauthorblockA{\textit{ETH Zürich}}
\and
\IEEEauthorblockN{Christelle Gloor \orcidlink{0000-0001-7031-2577}}
\IEEEauthorblockA{\textit{ETH Zürich}}
\and
\IEEEauthorblockN{Juan Angel García-Pardo \orcidlink{0009-0003-8813-1303}}
\IEEEauthorblockA{\textit{ETH Zürich}}
\and
\IEEEauthorblockN{Adrian Perrig \orcidlink{0000-0002-5280-5412}}
\IEEEauthorblockA{\textit{ETH Zürich}}
}

\begin{document}

\maketitle

\begin{table}[b]\footnotesize
  This document is an extended technical report of the work ``\fgpki{}: Securing Digital Assets Through(out) the Physical World'', \textit{Proceedings of the International Conference on Computing, Networking and Communications (ICNC)}, 2026. It includes additional details on the data structure, performance evaluation, location verification.
\end{table}

\begin{abstract}
  Although our lives are increasingly transitioning into the digital world, many digital assets still relate to objects or places in the physical world, e.g., websites of stores or restaurants, digital documents claiming property ownership, or digital identifiers encoded in QR codes for mobile payments in shops.
  Currently, users cannot securely associate digital assets with their related physical space, leading to problems such as fake brand stores, property fraud, and mobile payment scams.
  In many cases, the necessary information to protect digital assets exists, e.g., via contractual relationships and cadaster entries, but there is currently no uniform way of retrieving and verifying these documents.
  In this work, we propose the \fgpkifull{} (\fgpki{}), a geographical PKI that provides a global view of digital assets based on their geo-location and occupied space.
  \fgpki{} allows for the bidirectional translation of trust between the physical and digital world.
  Users can verify which assets are supposed to exist at their location, as well as verify which physical space is claimed by a digital entity.
  \fgpki{} supplements current PKI systems and can be used in addition to current systems when its properties are of value.
  We show the feasibility of efficiently storing millions of assets and serving cryptographic material based on precise location queries within \SI[detect-weight]{11}{\milli\second} at a rate of more than \num[detect-weight]{19000} queries per second on a single server.
\end{abstract}

\section{Introduction}
Public Key Infrastructures (PKIs) are critical for establishing secure communication between entities.
The web PKI, for instance, secures most web traffic by providing X.509 certificates,
vouched for by Certificate Authorities (CAs). Similarly,
the Resource PKI (RPKI) enables route origin validation for the Border Gateway Protocol (BGP).
However, PKIs face challenges, such as attacks on validation~\cite{BirgeLee2018,Dai2021},
mismanagement~\cite{sslmate2022}, and CA compromise~\cite{Comodo2011,Hoogstraaten2012}.
Improvements like certificate transparency (CT) enable retroactive attack detection by logging all issued certificates.

In high-security scenarios, PKIs should proactively prevent attacks.
This can be achieved by leveraging multiple sources of evidence,
similar to multi-factor authentication methods like OAuth~\cite{RFC6749} or FIDO~\cite{FIDO2022}.
Location is a natural choice, as it is intuitive for users, well-supported by mobile devices,
and rooted in societal practices such as land ownership and cadaster records.
Space ownership provides a tangible source of trust for enhancing PKIs.

We categorize the use cases enabled by location evidence into two classes defined by
the direction of the trust transfer.
In the \emph{first category}, trust rooted in the physical world---e.g., through a user's location service on their mobile device which can be additionally verified by the user by showing a map projection---is transferred to the digital world to retrieve digital assets linked to the location.
This category includes
adding location-anchors to a digital certificate
(e.g., users can obtain the postal service location-anchor if they visit a post office),
enumeration of present physical IoT devices and their owners
(e.g., used by delivery drones flying near user's premises),
and detection of fake payment services
(e.g., ``evil twin'' attacks where an adversary conceals their malicious service as the real one),
by enumerating all digital assets tied to the user's location, and thus revealing the trust conflict to the user.

In the \emph{second category}, trust rooted in the digital world,
in the form of a digital identifier---in particular a domain name---is transferred to the physical world.
This category includes
confirming location in the real world of hotels or rental properties
(e.g., via cadaster or a user-trusted accommodation platform),
binding trusted entities to IoT devices' physical presence
(e.g., enumeration of ATMs trusted by a subset of banks),
and protection of public and internationally recognized actors
(e.g., humanitarian organizations such as the ICRC).
This leads us to the following research question:
\emph{How can we extend the web PKI with geographical information to allow the
bidirectional transfer of trust between the physical and the digital world,
while preserving scalability?}

To address this, we propose the \fgpkifull{} (\fgpki{}) to allow owners of a physical space to define valid cryptographic material for their
claimed space using geo certificates (\geocerts{}). Users define trusted issuers locally,
enabling verification of spatial claims based on their territorial view, even during disputes.
The core component of \fgpki{}, the geo map server,
uses a sparse Merkle hash tree to efficiently handle overlapping \geocerts{} and support efficient queries at diverse granularity levels, i.e., both fine-grained queries and queries for large areas.
To improve scalability, data is stored at intermediate tree nodes.

We evaluate \fgpki{} by assessing the performance of certificate insertion and proof generation.
Our prototype can ingest an estimated 330 million globally-distributed \geocerts{} (see \cref{sec:evaluation:performance}), with a certificate renewal interval of 6 months, and serves \num{19000} queries per second, proving its readiness for Internet-sized deployments.
Client impact is minimal due to compact messages and lightweight computations.
Finally, we analyze the system's security and discuss approaches for location validation.

\section{Problem Statement}\label{sec:problem}

Public key infrastructures provide verifiable bindings between elements within a universe, e.g., domain names, and a set of properties, e.g., digital identities in the form of certificates.
In this work, we propose using an anchor in the physical world, i.e., relating to a physical space, providing an additional human-verifiable aspect for certificates.
In contrast to the domain namespace, physical space does not suffer from homograph attacks (due to uniform representation of space via longitude, latitude, and altitude) and physical space ownership is intuitive to users.
Henceforth, we refer to the user that validates a PKI binding as the \emph{relying party}.
By relying on an independent physical anchor, a relying party can increase the certainty of an existing web PKI binding.
A relying party can thus effectively transfer trust between the digital world, e.g., in the form of a digital certificate, and the physical world, e.g., by visiting a place and verifying the veracity of a claim in person.
This opens up several novel use cases.

\subsection{Use Cases}
To motivate and frame our work, and to convey a sense of usefulness to it, we present several real-world examples where transferring trust from the physical to the digital world, and vice versa, provides benefits to a relying party.

\myparagraph{Prevention of Payment Fraud}
Before authorizing a payment in a suspicious location, a relying party can check which digital entities claim this physical space, and then decide whether or not to proceed with the payment.
This detects ATMs set up by scammers, fake car parking meters, or legitimate mobile payment QR codes covered by fake codes, if the web PKI certificate presented during the transaction is not cryptographically tied to \geocert{} for that space.

\myparagraph{Detection of Fake Stores and Services}
A relying party wants to ensure that a store, e.g., selling products of a certain brand, or service, e.g., a WiFi hotspot, is legitimate.
Fake services, such as fake WiFi hotspots, known as evil twin attacks~\cite{Song2010a}, in a mall or airport, can be prevented by the space owner declaring the presence of the \emph{legitimate} hotspot, as \fgpki{} will show a discrepancy between the space owner's \geocert{} and the one presented by the WiFi hotspot.
Fake stores may also be detected by verifying whether the store at the relying party's location is legitimate (e.g., verified and approved by the city), which transfers trust from the physical to the digital world, disallowing e.g., payments to non-approved stores.
On the other hand, fake stores related to a certain brand can be detected by verifying whether the brand's \geocert{} claims to have a store at the relying party's location, and matching said \geocert{} to the one presented via the web PKI, i.e., transferring trust from the digital to the physical world.

\myparagraph{Protection of Public and Internationally Recognized Actors}
A relying party may want to check whether spaces allegedly belonging to public services, such as police stations, hospitals, vaccination clinics, etc., are legitimate before visiting them.
Similarly, a relying party may want to recognize whether a building is associated with a humanitarian organization such as the International Committee of the Red Cross (ICRC).
In both cases, the relying party's confidence in the service provider is improved and illegitimate providers can be avoided~\cite{CIAPakistan2021,Mukherjee2021}.

\myparagraph{Confirming Land Ownership}
A relying party may want to verify the veracity of land ownership (or of a temporary lease) before buying or renting apartments online to prevent real estate fraud.
Once the land is purchased or rented, the new owner can then claim the relevant space.

\myparagraph{Veracity of Location Statements}
In addition to real estate, the veracity of a location statement in the digital world is an important factor in reasoning about the represented physical assets.
This allows the relying party to verify proximity claims, e.g., ``the parking spot is close to the city center'', or the opposite, e.g., ``no noise from the highway reaches the hotel''.

\subsection{Requirements}

Although trust transfer from the physical to the digital world is missing in the current web PKI ecosystem, the inverse already takes place indirectly.
Businesses often add their address or a link to their location in a map service, such as google maps, to their website protected through TLS\@.
We argue that the current approach is insufficient to achieve strong guarantees for several reasons.
If an external map service is used, it necessitates another trusted entity.
If the address is directly added on the website, its detection and correct parsing is a non-trivial task.
Furthermore, the address may be modified or spoofed by an injected script, e.g., an external advertisement.
Additionally, in the absence of a verifying authority, the legitimacy of a location claim is only asserted by the domain owner itself.
Finally, since the website content is only protected with the symmetric TLS session key, it does not provide non-repudiation of the binding between domain and address.
For a system to support the aforementioned use cases, we identify the following main requirements:
\begin{itemize}
\item \textbf{Efficient Indexing}: Provide an efficient way of indexing location and space in the physical world at an adequate granularity for the use cases.
\item \textbf{Verifiable Bindings}: The bindings between physical space and digital identifiers must be verifiable by the relying party.
\item \textbf{Backward Compatibility}: Do not require changes to certificates that are currently in use, or devices deploying these certificates.
\item \textbf{Low Overhead}: Minimize impact on the current web PKI's performance and operational costs.
\end{itemize}

\subsection{Attacker Model}
We focus on a system to efficiently distribute space ownership statements and verify their veracity and completeness.
The attacker may: (1) read, block, and inject arbitrary messages between the relying party and  other entities in the system, and (2) compromise various actors participating in the system.
\Cref{sec:analysis} discusses the security implications of varying compromised actors.
We deem attacks on the location verification process, i.e., issuing space ownership statements, out of scope for this work, though we provide early observations on the problem space and mitigation strategies in the context of \fgpki{} in \cref{sec:location-verification}.
Furthermore, attacks on the location service of mobile devices are not considered, but there exists ample research on the prevention of such attacks~\cite{Brands1993,Singelee2005,Vora2006,Capkun2008,Yan2014,Nosouhi2020}.
Thus, the attacker cannot: (1) tamper with the location verification process, i.e., manipulate a location verification entity into issuing a false location ownership statement, (2) influence the perceived location of the relying party, and (3) compromise the relying party (as this would trivially defeat any authentication mechanism).

\subsection{Assumptions}\label{sec:problem:assumptions}
We assume that all entities, i.e., the relying party, space owner, and the infrastructure servers (geo log servers, map servers, and CAs) are bootstrapped with authentic certificates and public keys of their trusted entities.
For the log server, map server, and CA, the authenticity of these cryptographic objects can be verified through gossiping.
Additionally, we assume that log servers and map servers perform gossiping on their respective signed Merkle tree heads, to detect split view attacks.
For the relying party and space owner, the initial configuration containing relevant cryptographic objects should come from a trusted source, e.g., bundled in a browser, or served by a trusted organization.

\section{Design}
\fgpki{} is designed to satisfy the previous requirements, namely efficient indexing of physical space, cryptographically verifiable space-to-identifier bindings, backward compatibility, and low overhead.
This is achieved through space-related attributes stored in \emph{geo certificates (\geocerts{})}.
The owner of a space requests a \geocert{}, a cryptographic object representing the owner's claim of the space.
The relying party then validates the cryptographic object of interest, e.g., a TLS certificate, based on relevant \geocerts{}.

\subsection{Deployment Scenarios}
We present two deployment scenarios for \fgpki{}.
In an initial \emph{web PKI extension} deployment scenario, geographic information is piggy-backed on web PKI certificates as X.509 extensions and can thus reuse existing web PKI infrastructure.
This ensures that space owners can immediately enhance their existing web PKI certificates with geographic information and continue logging them with the existing CT infrastructure.
The only requirements are a web PKI CA that supports \fgpki{} and issues \geocerts{}, and a \fgpki{} map server.
The shortcomings of such a deployment are that the issuance of new geographic information is tied to the issuance of new web PKI certificates, reducing the flexibility of the system and putting additional burden on CAs and the logging infrastructure if geographic locations are frequently updated.

In a \emph{standalone} deployment scenario, we propose a separate certificate hierarchy based on certificate authorities dedicated to creating geographical certificates.
By decoupling the \fgpki{} hierarchy from the web PKI hierarchy, we ensure that they can be updated independently, enabling a wider range of use cases and simplifying the revocation process.
However, compared to the previous scenario, it additionally requires dedicated CAs and CT log servers.

In both deployment scenarios, \fgpki{} is incrementally deployable and relying parties immediately benefit from the additional protection by \fgpki{} even if there is only a single CA, map server, and log server.
As the partial deployment expands, relying parties can incorporate the newly participating CAs into their local trust view to extend the coverage of \geocerts{}, and increase \fgpki{}'s resilience by adding new map servers to their map server pool.
The concepts introduced in this section are orthogonal to the certificate hierarchy and thus apply to both deployment scenarios.
We will highlight the difference between the scenarios, where applicable.

\subsection{Components}
\begin{figure}
  \centering
  \includegraphics[width=0.6\linewidth]{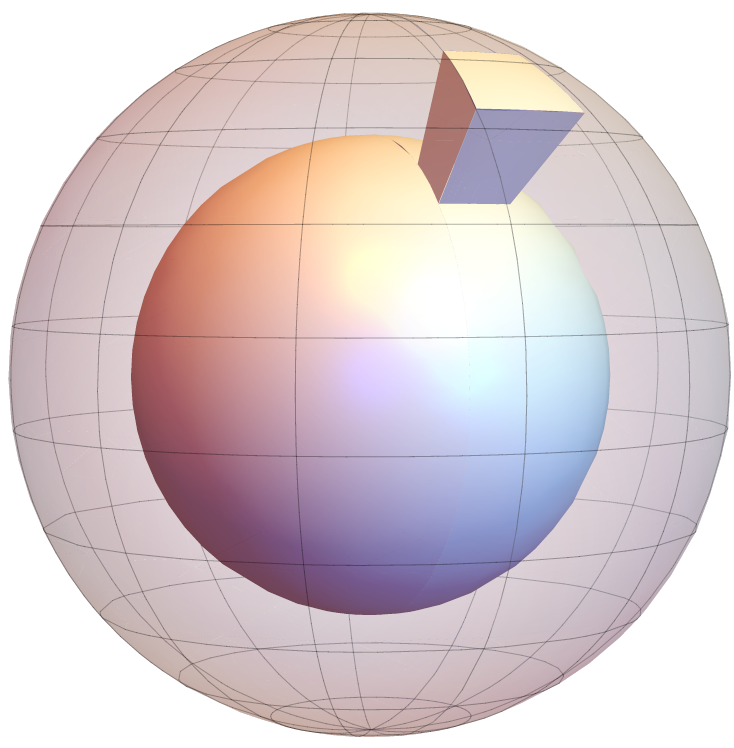}
  \caption{Visual representation of the space claimed by a \geocert{}.}\label{fig:frustum}
\end{figure}

\myparagraph{\geocerts{}}
\geocerts{} are used by the space owner to specify attributes for a claimed space.
Space is encoded as a set of frustums\footnote{A frustum is the portion of a solid that lies between two parallel planes cutting this solid.} as shown in \cref{fig:frustum}.
A frustum is defined by a closed polygon specified through latitude and longitude positions on the WGS84 ellipsoid approximation~\cite{wgs84} and a minimum and maximum altitude.
The volume of the frustum is the intersection of the ellipsoidal shell between the minimum and maximum altitude and the polygon extruded along the normals to the ellipsoid.

\begin{sloppypar}
The \geocert{} binds this space to attributes, such as the intended use of the certificate in this space, e.g., ``wireless access point'' or ``payment terminal''.
Another attribute is a URI identifying the space owner, which uses the custom scheme \texttt{\fgpkischeme{}} to represent a \fgpki{} URI\@.
This URI may include identifiers of the \geocert{} itself as path, query, or fragment within the URI\@.
For example, \texttt{\fgpkischeme{}://bank.com?branch=newyork} and \texttt{\fgpkischeme{}://bank.com?branch=london} could be used for two \geocerts{} of the New York and London branches of a bank, respectively.
The list of attributes is extensible, enabling unforeseen use cases.

Within a \webpkideployment{}, since \geocerts{} are encoded as X.509 extensions in existing certificates, they already support the creation of certificate chains.
However, in a \standalonedeployment{}, in addition to these attributes, \geocerts{} must contain fields to allow the creation of certificate chains similar to X.509 certificate chains, i.e., starting with a self-signed root certificate, possibly with multiple intermediate certificates, and ending with a leaf certificate.
An additional requirement for the validity of a \geocert{} chain is that the 3D-volume of a child certificate must be contained in the volume of its parent certificate.
\end{sloppypar}

\myparagraph{Geo CA}
\geocerts{} are issued by geographical certificate authorities (geo CAs).
Before issuing a \geocert{}, the geo CA validates the space claim of the certificate through a well-defined method, e.g., by physical inspection.
The validation method may be added as an attribute to the \geocert{}, as is discussed in \cref{sec:location-verification:security-considerations}.
In the \webpkideployment{} model, existing web PKI CAs act as geo CAs.
However, even in the \standalonedeployment{} model, existing web PKI CAs would likely also operate geo CAs, since they already have the necessary infrastructure to issue (geo) certificates.
Examples of standalone geo CAs are organization-specific CAs, e.g., an eduroam CA issuing \geocerts{} to protect wireless networks in universities, or in-person location verification CAs.
A concrete example of an in-person location verification CA is the Swiss postal company, which operates the Web PKI CA SwissSign and also performs address verification through their postal workers.
By collaborating with the national cadaster, SwissSign could offer their customers the automatic creation of \geocerts{} for their apartments and houses.

\myparagraph{Trust Preference}
A relying party defines its personal trust preference, which is a relative ordering of geo CAs, indicating the trust placed into each geo CA\@.
Hence, two relying parties can specify different, even conflicting, trust preferences and ensure that none of their less trusted CAs can claim space that is already claimed by their more trusted CAs.
Concretely, the trust preference consists of a set of $N$ geo CAs ($\text{CA}_i$) with their respective trust level ($\text{TL}_i$) and space ($V_i$) in which they are allowed to issue certificates, stored as the set of tuples $\text{TP} = \{\langle \text{CA}_i, V_i, \text{TL}_i\rangle\}_{i\in [1,N]}$.
This allows a relying party to accept less trustworthy CAs but restrict them to certain areas, e.g., use a state-operated CA of questionable trustworthiness but only for certificates issued within that state.
The concept of locally defined trust preferences was introduced in \fpkiorig{}~\cite{Chuat2022}, but instead of applying trust preferences on geo CAs based on the validated area, they evaluate the trustworthiness of web PKI CAs based on the validated domain.

\begin{sloppypar}
\myparagraph{Geo Log Server}
The geo log server collects certificates in a public append-only log.
It is identical to a certificate transparency (CT)~\cite{RFC6962} log server, except that it aggregates \geocerts{} instead of X.509 certificates.
In the \webpkideployment{} model, CT log servers remain unchanged.
\end{sloppypar}

\myparagraph{Map Server}
A map server periodically fetches \geocerts{} from log servers and stores them in a sparse Merkle hash tree to provide efficient proofs of presence or absence of certificates.
\fgpki{} uses two types of mappings, the geo map server which provides space-based lookup, i.e., returns all \geocerts{} claiming a certain space, and the domain map server which provides domain-based lookup, i.e., returns all \geocerts{} related to a certain domain.

\subsection{Workflow}
\begin{figure}
  % \includestandalone[width=\linewidth]{figures/overview}
  \includegraphics[width=\linewidth]{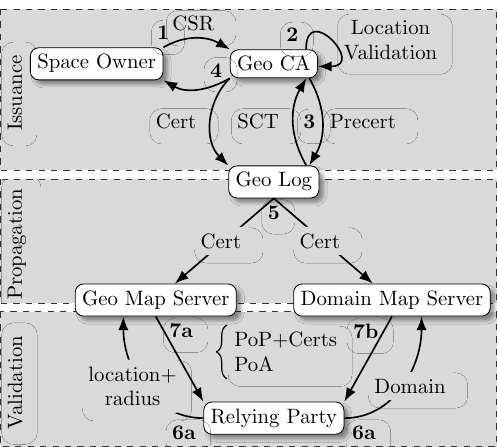}
  \caption{System Overview.}\label{fig:system-overview}
\end{figure}

\Cref{fig:system-overview} shows an overview of the interaction between the different \fgpki{} actors.

\myparagraph{\geocert{} Issuance} First, the space owner requests a \geocert{} from the geo CA with a certificate signing request (CSR) which includes all relevant certificate fields (step 1).
The CA validates the location claim in the CSR (step 2) and issues a precertificate.
The precertificate is sent to geo log servers, which reply with signed certificate timestamps (SCT) (step 3).
The CA encodes all SCTs and signs the final \geocert{}, which is sent back to the space owner and geo log servers (step 4).

\myparagraph{Propagating Certificates to Map Servers} The map servers collect \geocerts{} from the geo log servers and validate, via the MHT append-only property, that the received certificates are complete.
After collecting certificates for a fixed time interval (the maximum merge delay (MMD), which is typically \SI{24}{\hour}), the map servers ingest these certificates into their database (step 5) and release a new signed tree head (STH).

\myparagraph{Client-Side Validation} The relying party contacts, depending on the use case, either the geo or the domain map server to fetch \geocerts{}.
They send the requested location and the radius of interest to the geo map server and retrieve all existing certificates for a given location (step 6a), or they query the domain map server for all certificates issued to a given domain (step 6b).
The response from the map server is either (1) a proof of presence (PoP), i.e., a path to the relevant node in the Merkle tree including the set of certificates in this node, or (2) a proof of absence (PoA), i.e., a path in the Merkle tree leading to an empty subtree indicating the absence of any certificates (steps 7a and 7b).
Based on the certificates contained in the PoP or the PoA, the relying party validates the object of interest, e.g., by checking if a TLS certificate is issued to the same domain that claims this space.

\subsection{Data Structure Overview}
The map server enables relying parties to retrieve the set of all certificates located within a given space along with a proof of presence or absence.
To represent space, we use frustums described by a WGS84 polygon and a minimum and maximum EGM2008 altitude~\cite{wgs84}.

The main data structure is a geographically structured sparse Merkle tree (SMT).
More specifically, the SMT is a two-dimensional k-d tree dividing the earth's surface in a mosaic-like fashion.
Moreover, each node in the k\nobreakdash-d tree is the root of a binary altitude subtree that divides the altitude dimension.
The data structure thus effectively divides an ellipsoidal shell around the earth's surface.

\begin{figure}[t]
    \centering
    \includegraphics[width=\linewidth]{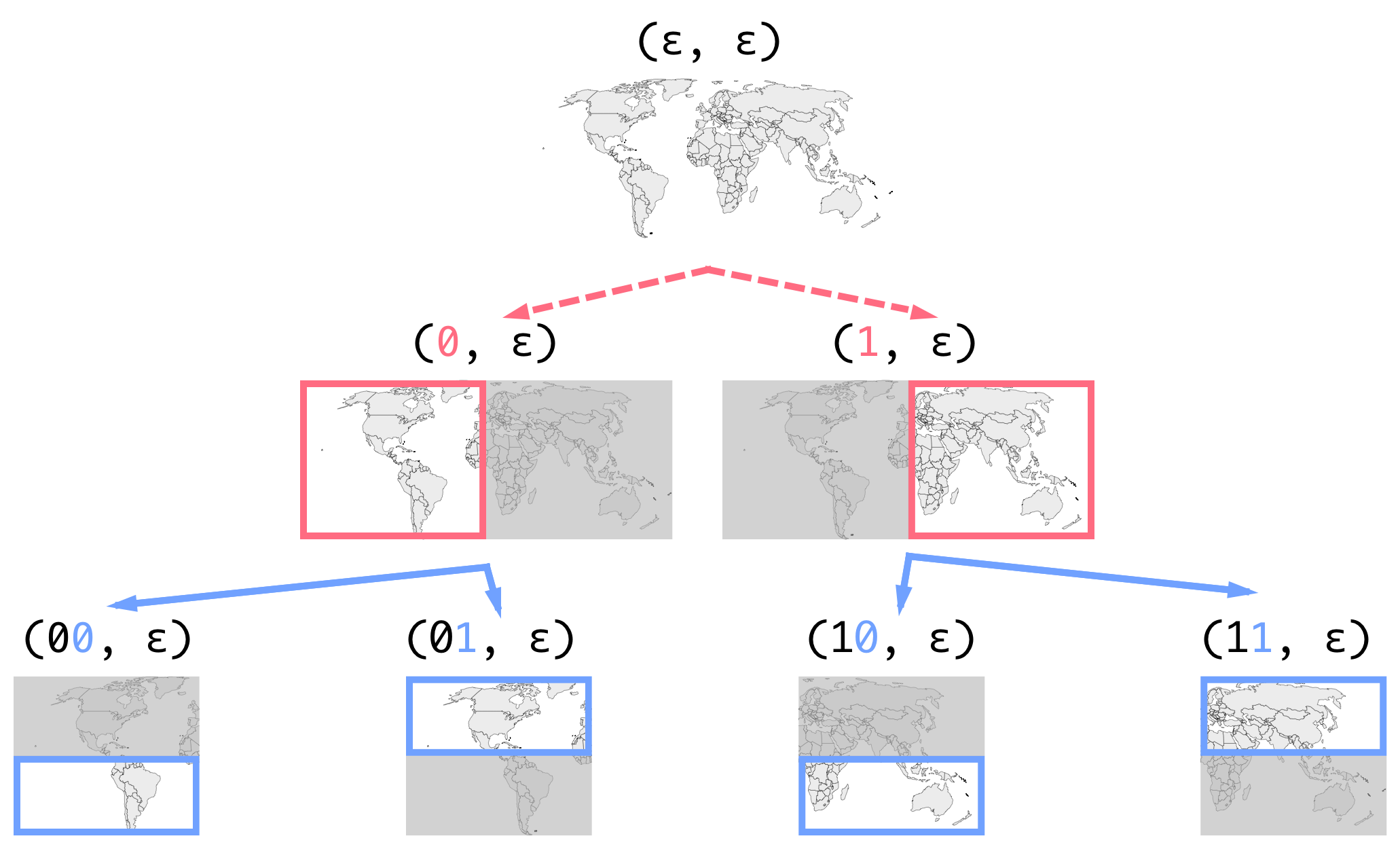}
    \caption{
    	The surface tree divides the world into smaller regions encoded by bit strings.
    	Arrows point from parent nodes to their children.
    	Red dashed arrows indicate longitude splits and blue solid arrows indicate latitude splits.
    }\label{fig:geopki-tree-2d}
\end{figure}

Nodes in the two-dimensional k-d tree represent the intersection of ellipsoidal sectors of the WGS84 ellipsoid approximation with the ellipsoidal shell.
The altitude subtrees rooted at these nodes then divide these frustums along the altitude dimension.
\Cref{fig:geopki-tree-2d} shows the two-dimensional k-d tree dividing the earth's surface.
Certificates can be associated with intermediate or leaf nodes in the tree.
This is particularly useful for certificates covering large volumes such as countries, states, or cities.

The SMT hash value of each node is computed by applying a hash function on the node's children hashes concatenated to the sequence of certificates associated with the node.
For leaf nodes, the hash is computed over the sequence of certificates.

\section{Data Structures}\label{sec:data-structures}
Analogous to certificate transparency, we aim to avoid the introduction of yet another trusted third party. 
Therefore, the data structure of the log and map servers must be auditable by any entity.

\newtheorem*{objectiveauditable}{Objective (Auditable)}
\greyborder{
  \begin{objectiveauditable}\label{objective:auditable}
    The data structure must be fully auditable by a third party to deter malicious behavior.
  \end{objectiveauditable}
}

A certificate must be authenticated before it is used to establish a secure connection.
Therefore, speed is imperative when authenticating certificates.
If the system introduces significant latency, it will not be adopted despite its security guarantees.
Thus, the primary objective of the data structures is to facilitate fast queries.
\newtheorem*{objectivefastqueries}{Objective (Fast Queries)}
\greyborder{
  \begin{objectivefastqueries}\label{objective:fast-queries}
    Given some volume, the data structures must be capable of efficiently retrieving the set of all certificates whose associated volumes intersect the query volumes along with a proof.
    In particular, it should allow for low latency, high throughput, and small response sizes.
  \end{objectivefastqueries}
}

To keep up with certificate issuance, i.e., ensure that all newly issued certificates can be added within a certain time limit, the system must facilitate efficient certificate insertion.
\newtheorem*{objectivefastingestion}{Objective (Efficient Updates)}
\greyborder{
  \begin{objectivefastingestion}\label{objective:fast-ingestion}
    The data structures must be capable of adding all newly issued certificates within a certain time limit.
  \end{objectivefastingestion}
}

Finally, given the constraints of the previous objectives, the storage overhead of the data structures should be low to limit the cost of log and map server operators.
\newtheorem*{objectivelowstorageoverhead}{Objective (Low Storage Overhead)}
\greyborder{
  \begin{objectivelowstorageoverhead}\label{objective:low-storage-overhead}
    The storage overhead incurred by the data structures should be minimized as long as this does not conflict with any other objective.
  \end{objectivelowstorageoverhead}
}

The data structure of the log server is analogous to an append-only public log server in CT,
and the data structure of the mapping from domain names to \geocerts{} is analogous to an \fpkiorig{} map server.
In the next two sections, we focus on the two main data structures of the volume-based map server: the sparse Merkle hash tree (SMT) for storing \geocerts{} and its consistency tree.

\subsection{Sparse Merkle Hash Tree}\label{sec:smt}
Inspired by \fpkiorig{}, an SMT is used to store the certificate data and prove the correctness of the responses.
Compared to \fpkiorig{}, instead of hierarchical trees following the domain-name structure, i.e., adding one SMT per subdomain, \fgpki{} makes use of a \emph{single} SMT with a geographical structure.
Each node represents a frustum inside an ellipsoidal shell around the earth and is encoded using a pair of bit strings.
The first entry in the bit string pair encodes an ellipsoidal sector (longitude and latitude), and the second an ellipsoidal shell (altitude).
The corresponding frustum is the intersection of the two.

\subsubsection{Surface Tree}
The first bit string encodes the position in the \emph{surface tree} shown in \cref{fig:geopki-tree-2d} where all nodes have corresponding surface areas.
Let the symbol $\epsilon$ denote the empty (bit) string.
Then the bit string pair $(\epsilon, \epsilon)$ is the root node of the tree.
A zero bit indicates that the path is traversing to the left, and a one bit indicates that it is traversing to the right in the tree, splitting the surface area associated with the parent node.
By specifying the full path from the root, there is a one-to-one correspondence between bit strings and nodes in the surface tree.
Hence, via the surface tree, each bit string encodes a unique surface area.
Moreover, each surface area, and therefore each node in the surface tree, has a unique corresponding ellipsoidal sector with respect to the WGS84 ellipsoid model.

\begin{figure}
    \centering
  		\centering
        \includegraphics[width=0.8\linewidth]{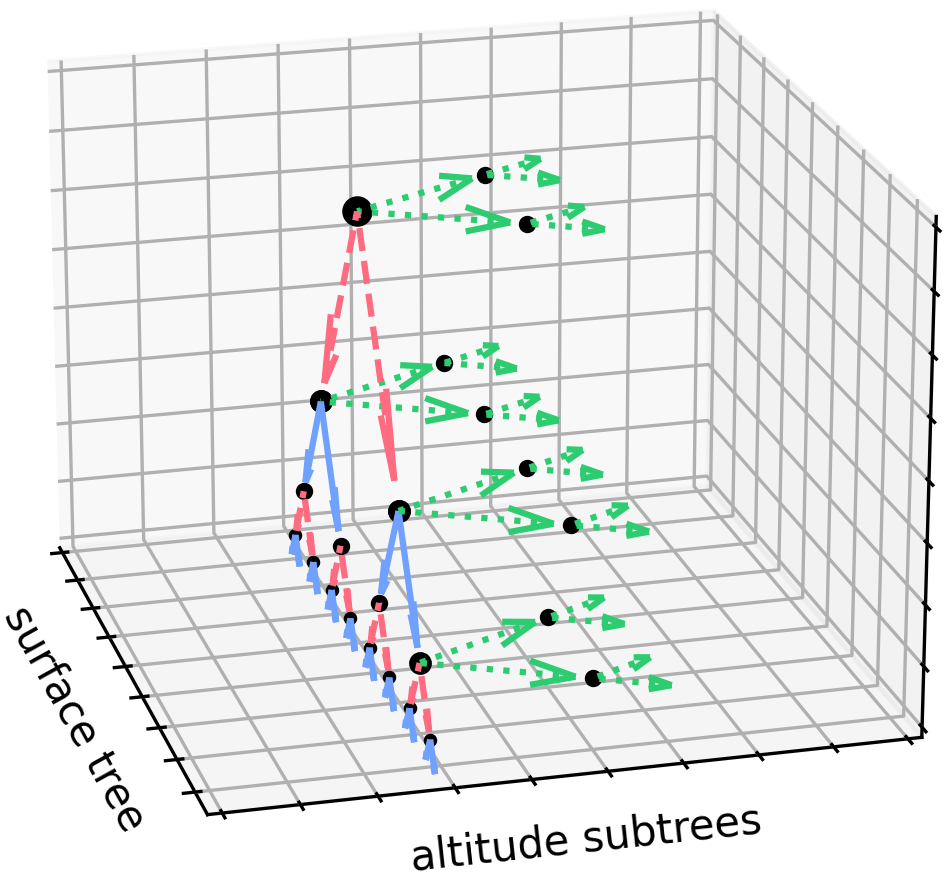}
        \caption{
    		Red dashed, blue solid, and green dotted arrows indicate longitude, latitude, and altitude splits, respectively.
    	}\label{fig:geopki-tree-3d}
\end{figure}

\subsubsection{Altitude Subtree}
The second bit string in the pair encodes the altitude bounds of the respective frustum, or rather the ellipsoidal shell, with respect to which the frustum is defined.
An empty string indicates the full ellipsoidal shell.
The bits in the altitude bit string split the default ellipsoidal shell along the altitude dimension, analogous to the surface tree.
In the tree structure, this is reflected by each surface tree node doubling as the root of a separate binary \emph{altitude subtree}, i.e., we first divide the two surface dimensions to obtain the target surface and then specify the altitude bounds of this surface.
The structure is visualized in \cref{fig:geopki-tree-3d} where the altitude subtrees are shown with green dotted arrows.

\subsubsection{Representation}
The height of the surface tree is fixed to a \num{51} bit encoding, and the height of each altitude subtree is fixed to \num{15} bits of encoding, achieving a precision of at least a meter in each dimension.

\begin{figure}
   \centering
   \includegraphics[width=0.8\linewidth]{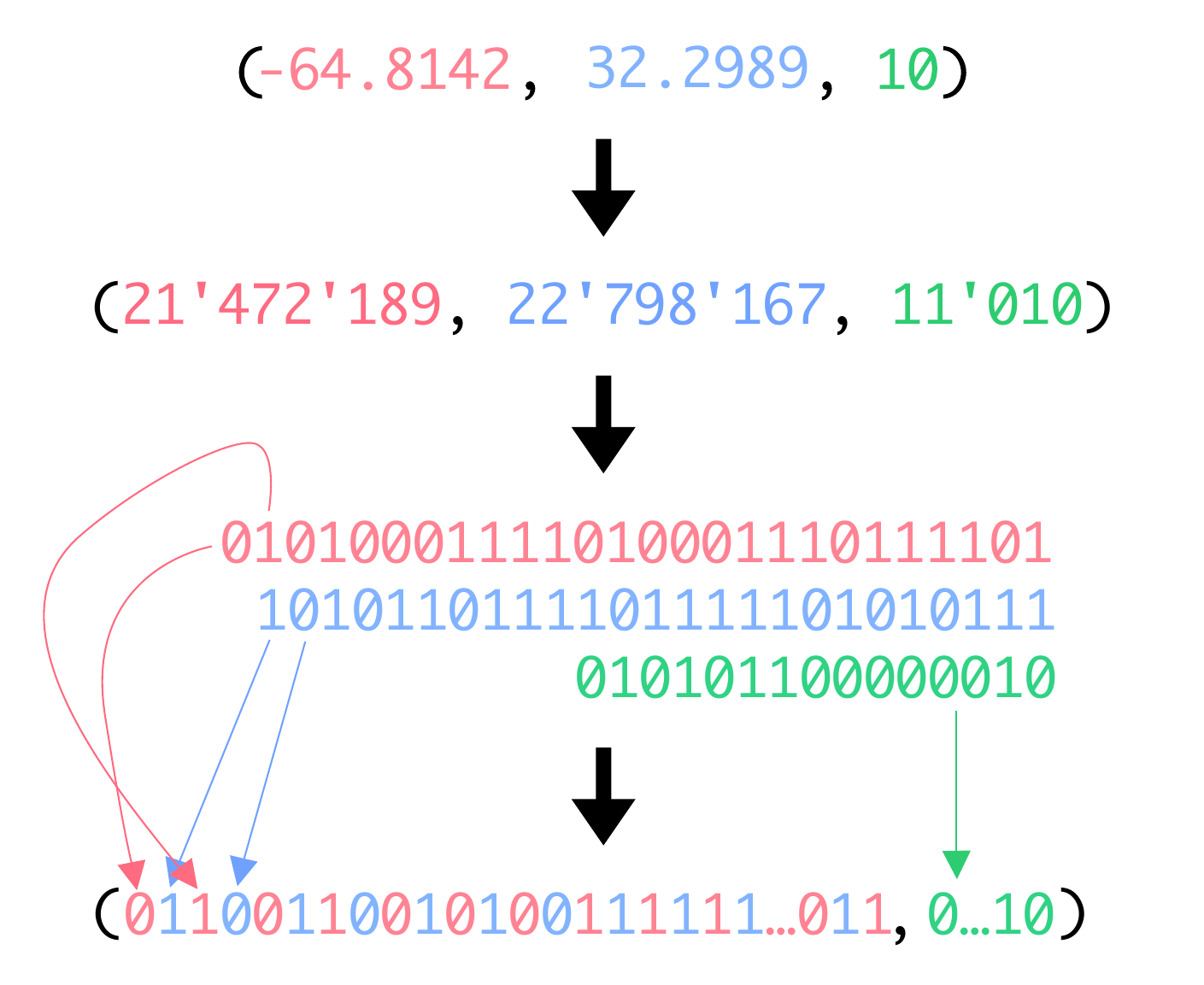}
   \caption{Discretization and Encoding of (longitude, latitude, altitude) tuples}\label{fig:coordinate-discretization}
\end{figure}

The first entry in the bit string pair encodes the $x$ (longitude) and $y$ (latitude) coordinates, and the second entry encodes the $z$ (altitude) coordinate as shown in \cref{fig:coordinate-discretization}.
Let \emph{surface bit string} denote the first entry of a pair and \emph{altitude bit string} the second.
The surface tree divides the earth's surface, such that a parent node covers the composition of its children's areas.
Each node and corresponding surface is encoded as a bit string, growing and becoming more specific (covering less area) the deeper the node's position in the tree.
The parent bit string is thus the prefix of its children's encoding.
Analogously to a k\nobreakdash-d tree, the divisions in the $x$ and $y$ dimensions are interleaved, starting with $x$.
Longitude values range from $-180$ to $180$ and latitude values from $-90$ to $90$.
For example, as visualized in \cref{fig:geopki-tree-2d}, the bit string pair $(0, \epsilon)$ encodes the area $[-180, 0) \times [-90, 90)$, the pair $(01, \epsilon)$ encodes $[-180, 0) \times [0, 90)$, and the pair $(010, \epsilon)$ encodes $[-180, -90) \times [0, 90)$.

Each node in the surface tree represents a frustum on the earth that extends from a fixed minimum altitude (\SI{11000}{\meter}) to a fixed maximum altitude (\SI{21767}{\meter}).
Nodes in the altitude subtrees divide the altitude dimension on each level.
For example, the bit string pair $(10, \epsilon)$ encodes the volume $[0, 180) \times [-90, 0) \times [-11'000, 21'768)$, $(10, 1)$ encodes $[0, 180) \times [-90, 0) \times [5'384, 21'768)$, etc.
While we expect many certificates to be altitude-specific (e.g., in cities and multi-story buildings), other certificates cover a larger area without such restriction (e.g., countries or national parks).
This design allows to only include altitude information where necessary, reducing the overall number of nodes.

The idea of interleaving coordinate bits is based on Geohash~\cite{Geohash,GeohashExplanation}.
The interleaving of the longitude and latitude bits results in a space-filling $Z$-order curve~\cite{Morton1966} which has the advantage of placing spatially nearby points close together in the SMT in the general case, increasing the number of common ancestors. 

\subsubsection{Hashes}
The tree division of the earth's volume is extended to a Sparse Merkle Hash Tree (SMT) by defining the computation of a hash value for each node in the tree based on its children and associated certificates.
For leaf nodes, the hash is derived exclusively from the certificates.
Concretely, the value is computed by applying a cryptographic hash function on a prepended constant, followed by the concatenation of certificate hashes associated with the leaf.
For all other nodes, the hash value is derived by applying the same cryptographic hash function on a prepended constant followed by the concatenation of the node's children hashes and the concatenation of the certificate hashes.
In both cases, the sequence of certificate hashes is sorted lexicographically.
Since the Merkle hash tree is sparse, the majority of logical children do not exist, and their hash is replaced by a default value.

\subsubsection{Points to SMT Nodes}
A point tuple $(x,y,z)$ at longitude $x$, latitude $y$, and altitude $z$ is mapped to discretized unsigned integer coordinates.
The maximum discretized integer coordinate for the respective dimension is set to result in a precision of at least one meter.
The bits of the longitude and latitude integers are then interleaved to obtain the first entry in the bit string pair, while the second entry directly corresponds to the bits of the integer representing the altitude.

\subsubsection{Volumes to SMT Nodes}
While mapping arbitrary spatial points to SMT nodes is trivial, mapping complex volumes is not.
Certificate volumes may be arbitrary in shape and size and rarely correspond to the volume of a single SMT node.
Moreover, queries for volumes, such as spheres around a location, must be mapped to a set of bit string pairs referencing the corresponding SMT nodes.

In both cases, the volume $V$ is approximated using a set of frustums.
For the assignment of certificates to SMT nodes, only over-approximations of the volumes achieve correctness, and a client may do an additional filtering step if too many certificates are returned.
Under-approximations may result in omitted certificates, which cannot be corrected for.
Therefore, we over-approximate the volumes.

A trivial over-approximation computes the deepest SMT node whose frustum encompasses all of $V$.
Unfortunately, this approach assigns all certificates' volumes intersecting the prime meridian to the root node or its altitude subtree.
As a result, these certificates must be included in the response to any query.
Even worse, all queries whose volume intersects the prime meridian will result in a query for the entire surface of the earth.
Instead, we compute a set of disjoint bit string pairs associated with smaller frustums based on the following heuristic:
Let $S$ be the two dimensional surface projection of $V$.
We first compute a set of surface bit strings whose union covers $S$ and then a set of altitude bit strings whose union covers the full altitude range of $V$.
The resulting bit string pair approximation corresponds to the cross product of the two sets, i.e., a set of frustums covering $V$.

We first determine an appropriate grid size, represented by a surface tree depth, and then compute the set of grid cells intersecting $S$.
We use an additional parameter for the derivation of the grid size, the relative grid size $f \in [0, \infty)$.
More specifically, we compute the surface tree depth such that the grid cell's corresponding surface area is equal to $f$ times the area of $S$.
If the factor $f$ cannot be achieved exactly, we choose the smallest grid size which covers at most $f$ times the area of $S$.
Note that if $f$ times the area of $S$ is smaller than the smallest surface at this location, the surface tree depth is set to the maximum value.
The surface defined by the grid will rarely completely cover $S$ because $S$ is, in general, not aligned to the surface tree split lines.

\begin{figure}
    \centering
    \begin{subfigure}[t]{0.18\textwidth}
        \includegraphics[width=\textwidth]{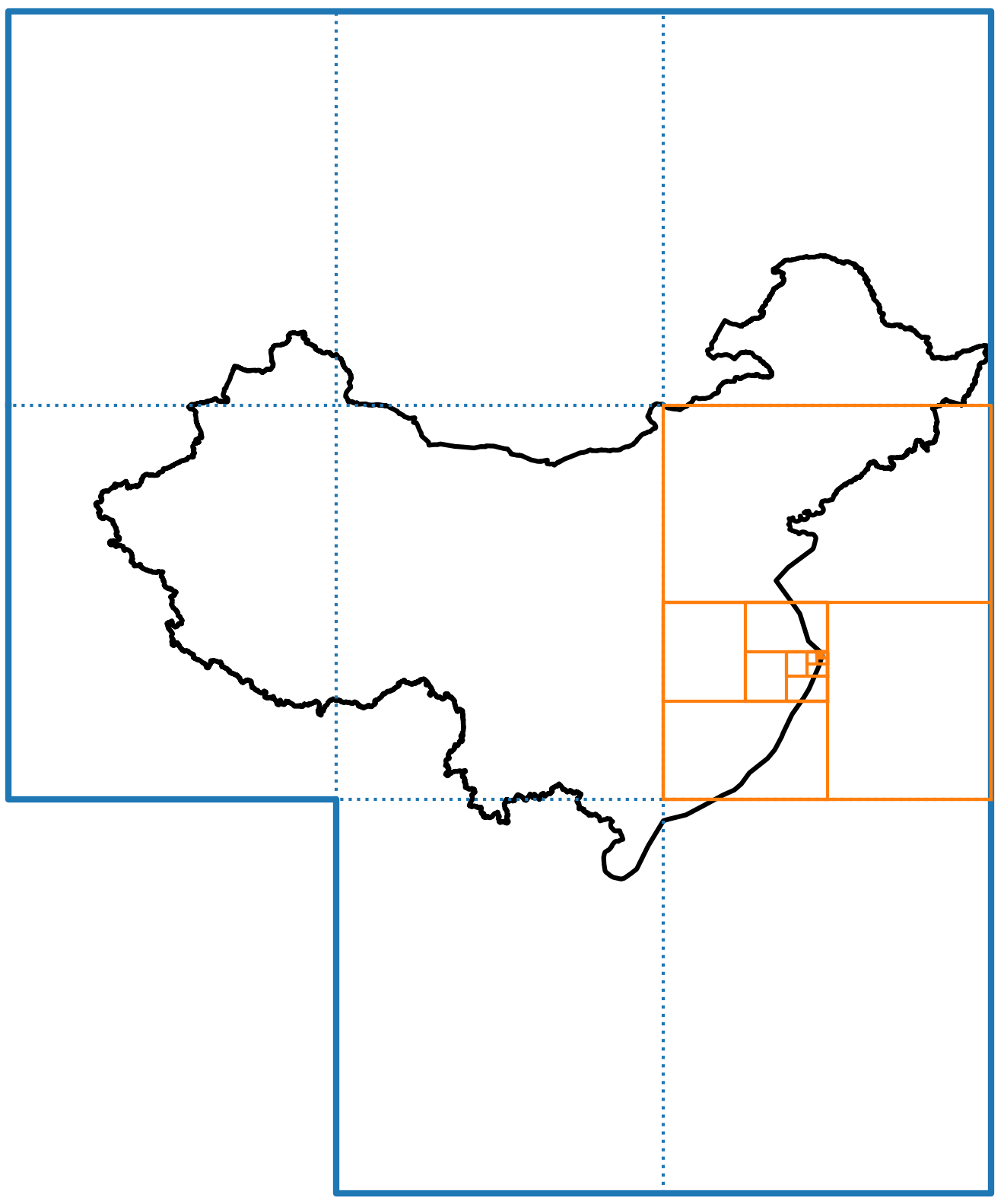}
        \caption{
        	$f=$ \num{1}
        }
    \end{subfigure}
    \hfill
    \begin{subfigure}[t]{0.28\textwidth}
        \includegraphics[width=\textwidth]{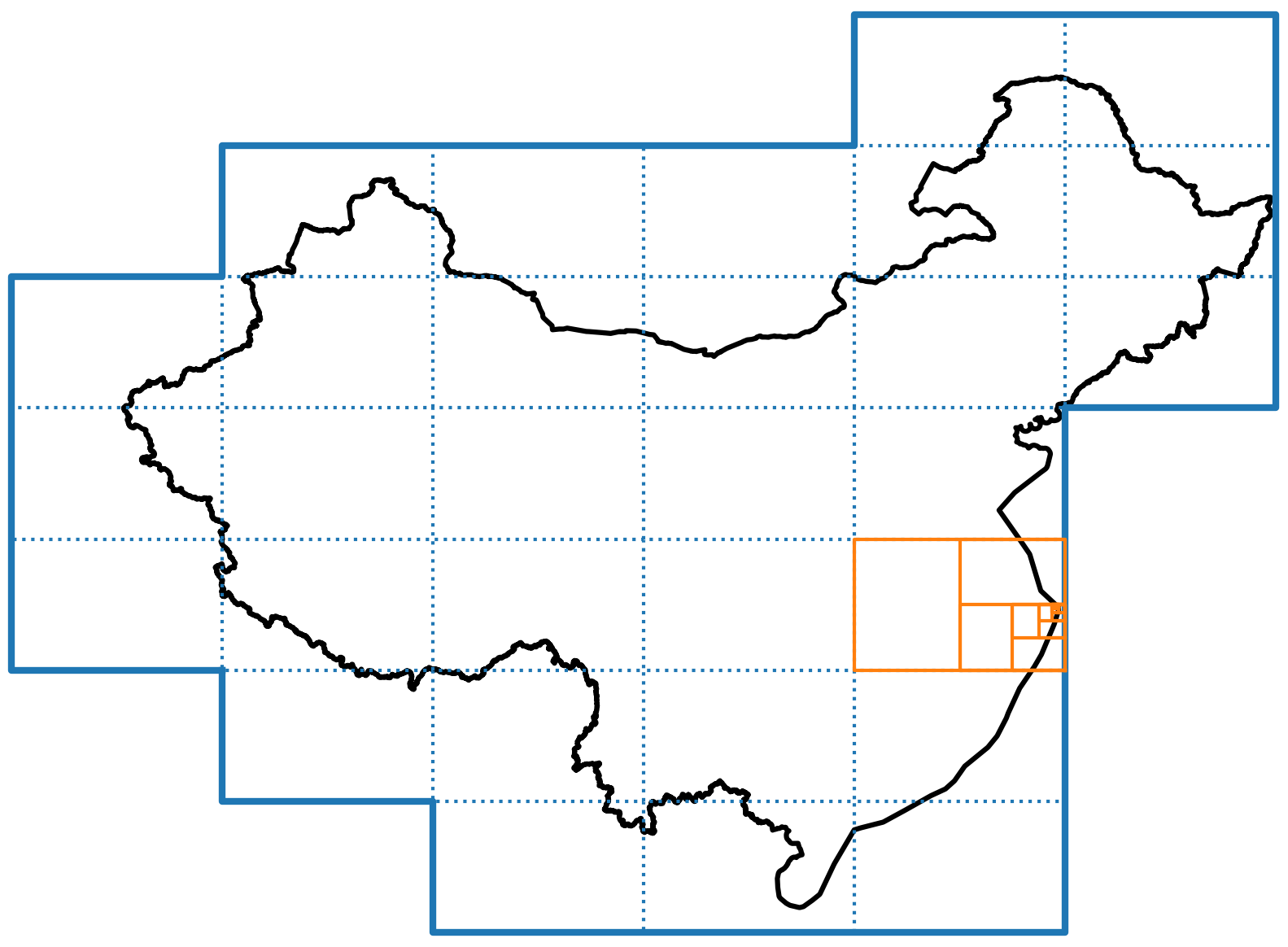}
        \caption{
        	$f=$ \num{0.1}
        }
    \end{subfigure}
    \caption{
    	Visualization of the assignment of SMT nodes to a set of input polygons for different relative grid sizes $f$.
    	The smaller orange rectangles show the ancestors that were traversed until arriving at the desired grid size. 
    }\label{fig:geopki-voxel-assignment}
\end{figure}

We compute the bit string of a fixed point on the polygon as described before.
We then traverse the ancestors until we arrive at the depth corresponding to the desired grid size.
Then a breadth-first search among spatial neighbors (not tree neighbors) finds the set of all surface tree nodes whose surfaces intersect $S$.
Their union covers $S$.
Any sibling nodes in the set are replaced by their parent to reduce the number of bit string pairs, while maintaining coverage.
The algorithm is visualized in \cref{fig:geopki-voxel-assignment}.

For the altitude bit strings, an analogous process could be performed.
However, we instead compute the single, longest altitude bit string (covering the smallest possible range) including the minimum and the maximum altitude.
This corresponds to the longest common prefix of the two bit strings encoding the minimum and maximum altitude.
This approach suffers from the aforementioned issue that intersections with the first split line force the result to be the empty bit string representing the full altitude range.
The advantage is that the result is a single bit string whose cross product with the set of surface bit strings has the same cardinality as the set of surface bit strings.
This minimizes the number of SMT nodes a certificate is assigned to and therefore the database overhead.

\subsubsection{Proofs of Completeness}
Given a query for a volume, the map server must prove that the returned data is part of the SMT and that no certificate was omitted.
Let $N$ be the set of non-empty nodes whose corresponding volumes intersect the query volume.
For each such $n \in N$, the server must send the set of all certificate hashes to the client to allow the node's hash to be recomputed.

By definition, parents always cover a super-volume of their children, and therefore $N$ includes the transitive closure of the parent relation.
For nodes with one child not in $N$, this child's hash has to be included in the server's response because it is also input to the node's hash computation.

\subsection{Consistency Tree}
In PKIs, \emph{consistency} usually refers to the adherence to a protocol for modifying an auditable data structure.
In CT, consistency between any two versions of a CT log can be proven through its append-only property.
There is no comparable method to prove consistency between different versions of an SMT as described in this work.
A solution proposed by \fpkiorig{} to ensure consistency between different SMT versions is to use a second (non-sparse) tree in addition to the SMT called a \emph{consistency tree}.
\fgpki{} leverages such a consistency tree that records the version history of the map server's SMT in the form of a chronological list of signed map server tree heads (SMH).
An SMH is a structure that contains the root hash of the SMT, a timestamp, a set of CT logs from which the map server derives its content, and a signature calculated over these fields.
Each CT log is described by the tuple (log id, STH), where STH = (timestamp, tree size, tree root).
The map server must ensure that for each listed CT log with a tree size $N$, the first $N$ entries in this CT log were used to create the given SMT.
The signature makes this claim non-repudiable and verifiable by auditors.
Concretely, a missing certificate from a CT log that should have been included according to the tree size is a proof of misbehavior.

The consistency tree has the same properties as a CT log append-only tree and requires the root node of the consistency tree and the size of the tree to be signed and published as the \emph{signed consistency head}.
Support for proofs of inclusion then enables the detection of split-view attacks by observers if the signed consistency heads are gossiped among them.
Although the consistency tree records the temporal history of versions, it does not guarantee that changes between versions are correct and follow the specification.
Instead, the CT log tuples provide auditors with all information necessary to reconstruct a replica of any past SMT by querying the respective CT logs.
The disadvantage of this approach is that it puts a heavy burden on auditors, since verifying the consistency requires reconstructing an SMT potentially consisting of hundreds of millions of entries.

\subsection{Revocation}
Revocation of \geocerts{} varies with the deployment model.
With the \webpkideployment{}, existing web PKI revocation mechanisms, such as OCSP and CRLs, are used.
In the \standalonedeployment{}, revocations are included in the log and map servers, allowing the relying party to simultaneously fetch \geocerts{} and revocations.
Hence, relying parties are always provided with an up-to-date view on all revoked \geocerts{}.

\section{Implementation and Evaluation}
We implemented both the geographical map server and a client using Go. % 1.20.5
The map server stores the data in a PostgreSQL database in three tables, one table for the \geocerts{}, and two tables for the SMT. % v14
One SMT table serves queries, whereas the other ingests new data.
To issue a new signed map head (SMH), which represents the state of the SMT at a given point in time, the two tables are swapped.
For the assignment of certificates to SMT nodes, we use a relative grid size of $f=\num{0.1}$.
The client approximates the query circle with a relative grid size of $f=\num{1}$.
To evaluate \fgpki{}, we derive a dataset based on real-world data, estimate the density of \geocerts{}, and finally measure the performance of our implementation.
The source code and evaluation results are publicly available~\cite{gecko-source-code}.

\subsection{Methodology}
\begin{sloppypar}
Our evaluation focuses on the geography-based map server, \fgpki{}'s core data structure, and the communication between the relying party and the map server.
On the map server, we focus on the \emph{ingestion} (the rate with which new \geocerts{} are added) and the \emph{throughput} with which responses (including proofs of presence or absence) are generated.
Regarding communication, we focus on the overhead in terms of bandwidth, i.e., map server \emph{response sizes}, and latency, i.e., map server \emph{response time}.
\end{sloppypar}

\begin{sloppypar}
The map server's throughput may be scaled up by replication, e.g., load balancing among multiple map servers or replicating databases.
Similarly, the map server's ingestion may scale up by sharding, e.g., storing \geocerts{} in different map servers based on their location, usage, or domain namespace.
Hence, the most relevant metrics are the response size and the response time which cannot easily be improved.
\end{sloppypar}

\begin{figure}[t]
  \centering
  \includegraphics[width=1\linewidth]{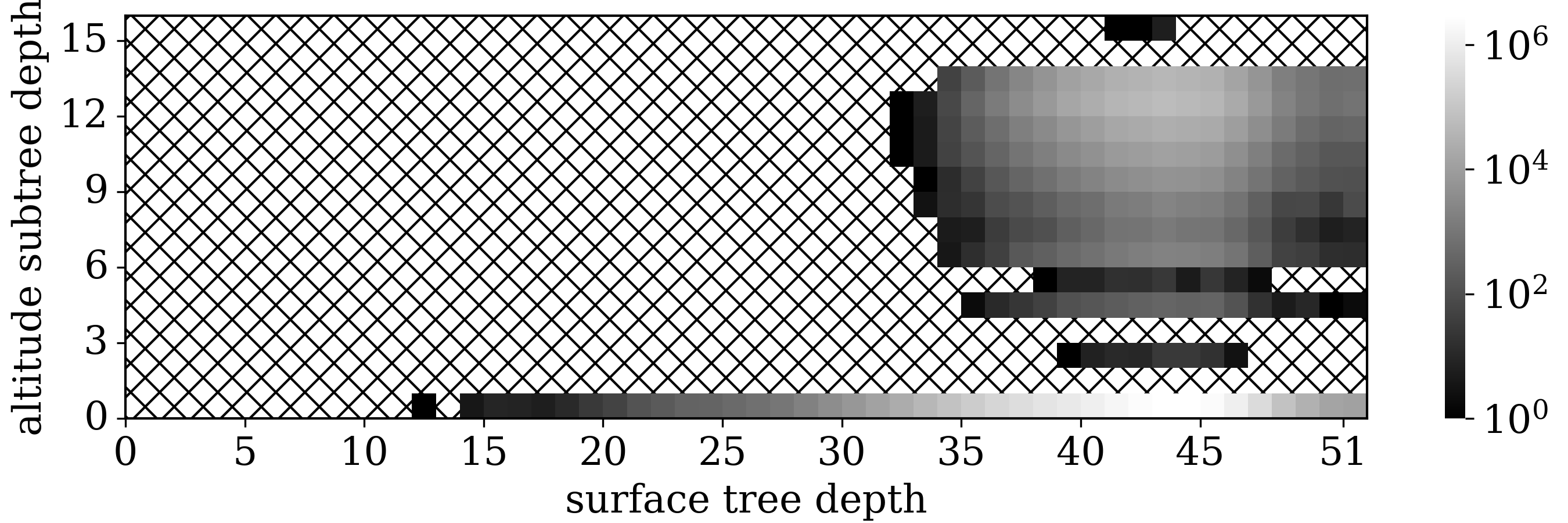}
  \caption{
    Heatmap of the tree depth distribution of sparse nodes.
    The shade indicates the number of nodes.
    Note that the color map is scaled logarithmically.
  }\label{fig:evaluation:leaf-distribution}
\end{figure}

\subsection{Datasets}\label{sec:evaluation:dataset}
We primarily derive our dataset based on Open Street Map (OSM) by extracting all location data associated with websites.
We chose OSM, instead of the more complete Google Maps, due to its powerful API capabilities and since it is freely available to researchers.
If only point locations (and no polygon shapes) are available, we correlate the points with the buildings covering them, using Delaunay triangulation to assign space if multiple points are located in the same building.
We discard \geocerts{} with excessively large payloads, i.e., we exclude payloads strictly greater than \SI{3.25}{\kibi\byte}, (99th percentile).
In practice, these payloads would be replaced with less precise polygons with reasonable payload sizes.
Note that this does not affect the security of the system as long as the less precise polygons completely cover the real polygon.
If OSM floor data is available, we use the digital elevation map STGTM v003~\cite{Nasa2019} and a floor height of \SI{3}{\meter} to approximate the frustum heights.
The final dataset consists of \num{1706375} \geocerts{}.
\Cref{fig:evaluation:leaf-distribution} shows the distribution of the SMT tree depths of all sparse SMT nodes in our dataset.
Sparse nodes are nodes whose descendants do not have any associated data, see \cref{eq:sparse-predicate:leafs,eq:sparse-predicate:intermediate} in \cref{app:smt-formalized}.
We can see that the vast majority of sparse nodes have a tree depth of around 43 and do not have any floor data.

To estimate an upper bound on the total number of expected \geocerts{}, we combine OSM data with the GHS-POP - R2022A population dataset~\cite{Schiavina2022,Freire2016}, and the more complete Google Maps database.
Since the API of Google Maps is capped at 60 results per query and not freely available, we use random sampling in combination with the Bayes' theorem to estimate that a total of \num{330} million websites may acquire \geocerts{}.
We want to emphasize that this is a crude estimate based on current data, and the future situation may turn out to be vastly different.

Additionally, because the SMT is structured by geography, i.e., not balanced, the tree depth, and therefore response sizes, only depend on the local density of certificates.

\begin{figure}[t]
	\centering
    \includegraphics[width=1\linewidth]{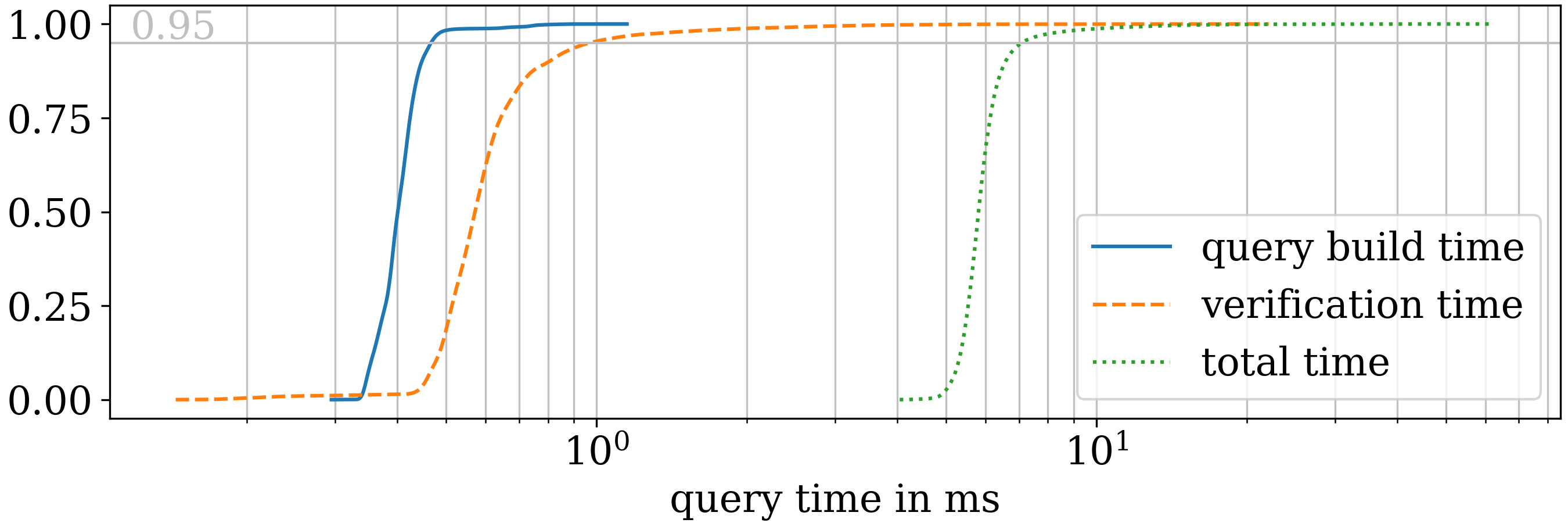}

    \caption{
    	CDFs of different time measurements.
    	Verification time is the time it takes the client to locally verify the server's response.
    	Total time includes the bit string computation time, query latency and verification time.
    	The horizontal line indicates the 95th percentile.
    }\label{fig:evaluation:client-performance:time}
\end{figure}

\begin{figure}[t]
	\centering
    \includegraphics[width=1\linewidth]{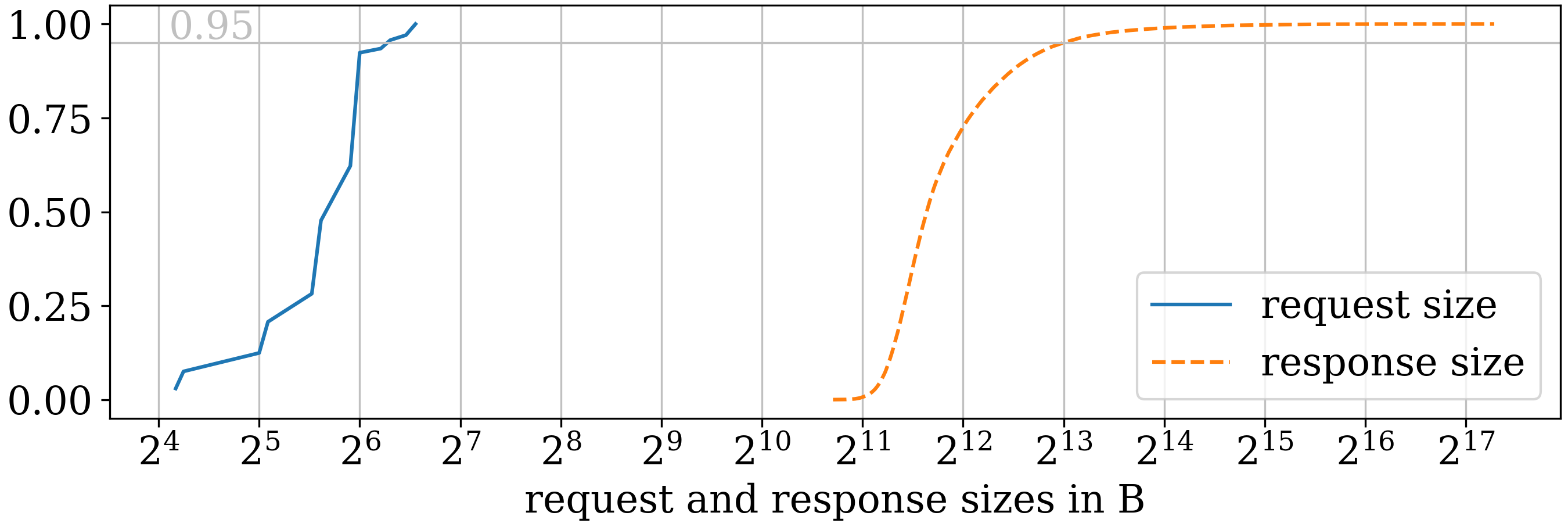}
    \caption{
    	CDFs of request and response sizes.
    	The horizontal line indicates the 95th percentile.
    }\label{fig:evaluation:client-performance:request-response}
\end{figure}

\subsection{Setup}

We evaluate the performance using a single machine with an Intel(R) Xeon(R) Gold 6242 CPU @ 2.80GHz (64 cores), 190GB of RAM, and an SSD, running Ubuntu 22.04.2 LTS (GNU/Linux 5.15.0-60-generic x86\_64), showing that we can support a large number of clients and millions of \geocerts{} with limited resources.

For throughput and latency measurements, we pre-generate a shuffled list of \num{1706375} locations to query, one point for each certificate in the dataset.
Each query requests a surface radius of \SI{10}{\meter} and includes certificates at any altitude to represent the worst case where the client does not have reliable altitude measurements.

For the ingestion measurements, we randomly sample locations for new certificates within Europe, according to a low resolution website density distribution based on OSM data.
Within each grid cell of the dataset, the query location is sampled uniformly at random and a certificate associated with a $\qty{10}{\meter}\times\qty{10}{\meter}\times\qty{3}{\meter}$ cube is ingested at a uniformly random base altitude..
The input thus follows the existing distribution but is still very likely to create new SMT nodes because of the low resolution of the distribution.

\subsection{Performance}\label{sec:evaluation:performance}

\Cref{fig:evaluation:client-performance:time} shows a client processing overhead on the order of milliseconds which is negligible compared to the HTTPS connection establishment overhead.
The measured query latency is below \SI[round-mode=places,round-precision=1]{7.096}{\milli\second} (95th percentile), excluding network delays.
We deem this acceptable, especially in light that a query can launch simultaneously to the network request for the resource (e.g., the DNS query resolving the domain of a webpage).

\Cref{fig:evaluation:client-performance:request-response} shows that, while the measured request sizes with a maximum of \SI{94}{\byte} are negligible, response sizes on the order of several kilobytes can cause delays, depending on the available throughput.
However, even under the assumption of a 3G network with limited throughput of \SI[per-mode=symbol]{384}{\kilo\bit\per\second}, the latency overhead of the response transmission is still below \SI{171}{\milli\second} (95th percentile).

\begin{figure}[t]
  \centering
  \includegraphics[width=1\linewidth]{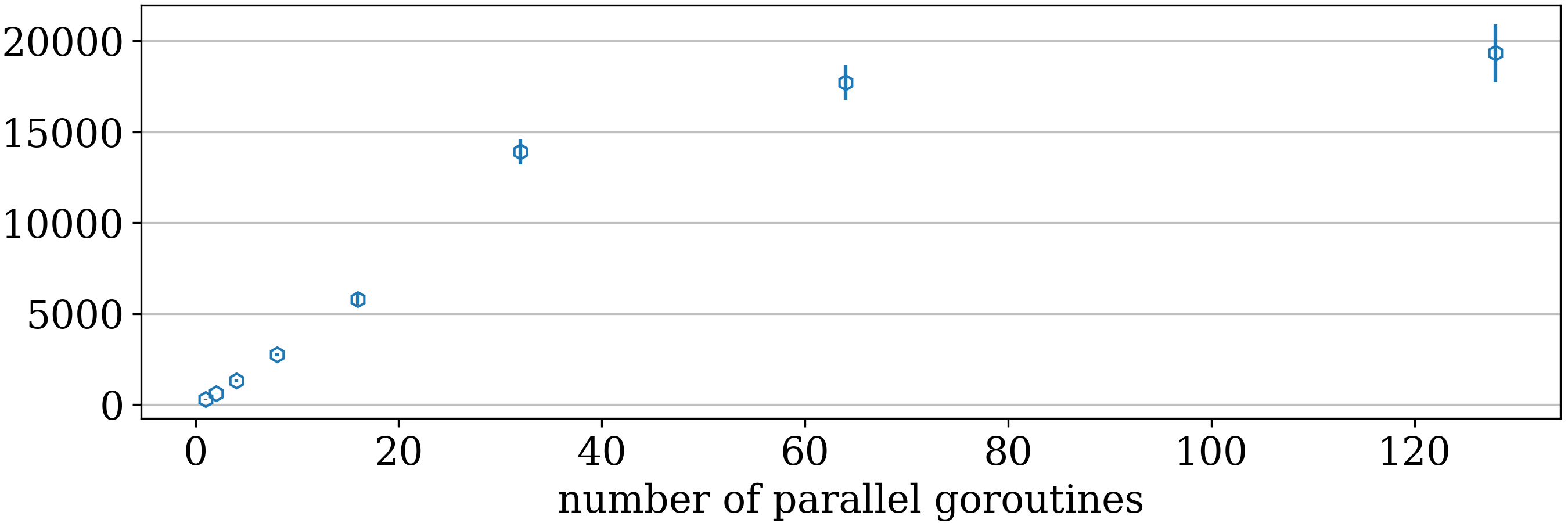}
  \caption{
    Map server throughput in terms of queries per second based on the maximum number of parallel goroutines.
  }\label{fig:evaluation:server-performance:throughput}
\end{figure}

\begin{figure}[t]
  \centering
  \includegraphics[width=1\linewidth]{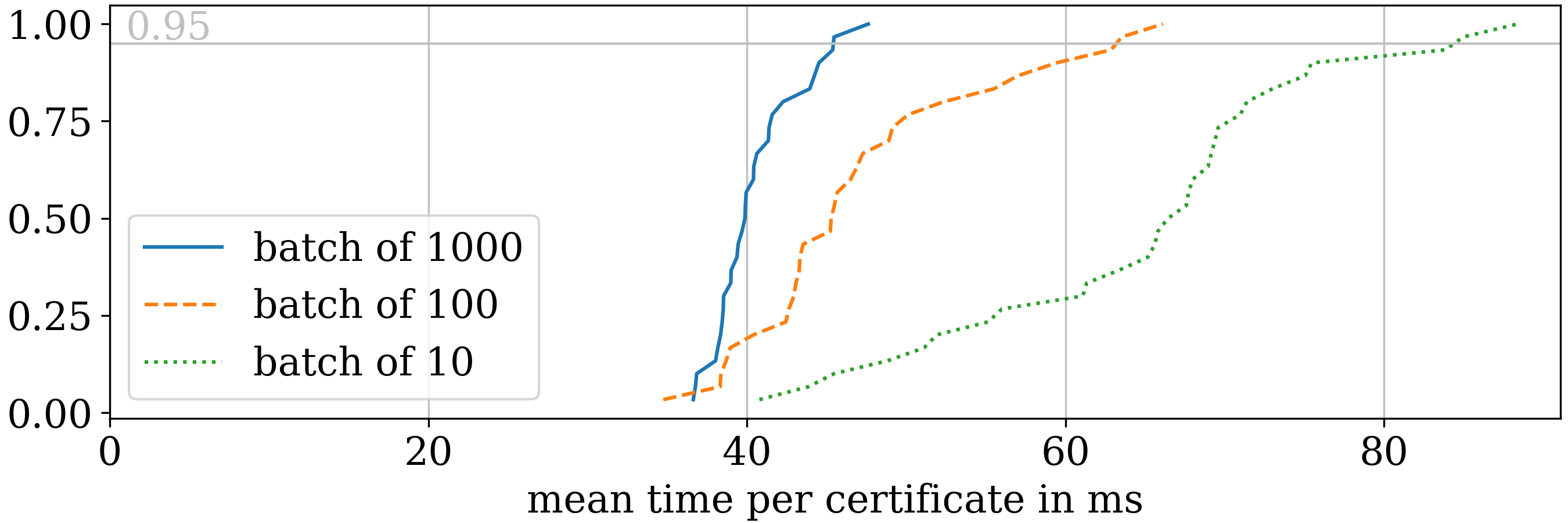}
  \caption{
    Map server ingestion rate in terms of mean time per certificate ingestion for different batch sizes.
  }\label{fig:evaluation:server-performance:ingestion}
\end{figure}

We evaluate \fgpki{}'s scalability by throughput and ingestion measurements.
A single machine handles up to \num{19000} queries per second, as shown in \cref{fig:evaluation:server-performance:throughput}, and takes a mean time of around \SI{40}{\milli\second} to ingest a single certificate for batch sizes of 1000, as shown in \cref{fig:evaluation:server-performance:ingestion}.
Based on our previous estimate, we could support the ingestion of our estimated upper bound of \geocerts{} (\num{330} million) with a certificate lifetime of roughly half a year.
We cannot predict the number of \geocerts{} that \fgpki{} must support in the future. The emergence of novel use cases may require one \geocert{} per person or IoT device.
However, the system can be easily scaled to multiple machines via sharding, load balancing among map servers, and database replication.

\section{Security Analysis}\label{sec:analysis}
In this section, we analyze the security of \fgpki{}.
In particular, the security properties of the verifiable space-to-identifier bindings used by the relying party to validate the cryptographic object of interest.
Note that in some cases, the \geocert{} itself is the object of interest, especially in the \webpkideployment{} model.

\fgpki{} has two goals: (1) fetching all relevant \geocerts{} and (2) verifying the object of interest once all relevant \geocerts{} are available.
We split the first goal into two objectives with distinct attacker models, namely the \emph{detection} and \emph{prevention} of attacks.
As described in \cref{sec:problem:assumptions}, we assume that the relying party is bootstrapped with the necessary cryptographic material, in particular authentic certificates and public keys of all trusted CAs, log servers, and map servers.

The first objective is to detect attacks by the different actors, namely the CAs that issue \geocerts{}, the log servers that operate an append-only \geocert{} log, and the map servers that aggregate \geocerts{} by space and domain, respectively.
\newtheorem*{objectivefetchdetect}{Objective (Fetching---Attack Detection)}
\greyborder{
  \begin{objectivefetchdetect}\label{objective:fetch-detect}
    A relying party that receives an authentic response from a map server can detect an attack by a CA, a log server, or a map server (or any combination of colluding entities) after the attack has occurred and prove the occurrence to other entities.
  \end{objectivefetchdetect}
}
This objective holds since all operations produce cryptographic evidence, i.e., \geocerts{} issued by CAs and signed MHT roots for log and map servers, which can be used to pinpoint the misbehaving entity.
The detection of split-view attacks by one or more log servers or map servers is discussed in more detail at the end of this section.

However, ideally, attacks should not only be detected after a (possibly lengthy) time period, but instead be prevented.
Analogous to \fpkiorig{}, the relying party requests responses from multiple trusted map servers and considers the union of all certificates from all map server responses to reduce the impact of a single malicious map server.
Hence, the second objective is preventing attacks by any number of possibly colluding actors, as long as a single benign map server is reachable and the certificate was logged by a log server.

\newtheorem*{objectivefetchprevent}{Objective (Fetching---Attack Prevention)}
\greyborder{
  \begin{objectivefetchprevent}\label{objective:fetch-prevent}
    A relying party that receives an authentic response from a benign map server is ensured that all \geocerts{} for the queried space are returned if each certificate is covered by at least one benign log server.
  \end{objectivefetchprevent}
}
This objective holds since the responses from all map servers are combined via union, hence a certificate from the benign map server's response cannot be removed.

Furthermore, as long as each certificate is covered by at least one benign log server, the log server adds the certificate within the maximum merge delay, e.g., \SI{24}{\hour}, after the respective SCT was issued, and the map server ensures that \emph{all} certificates from all log servers are fetched.
The guarantees of this objective are thus directly related to the number of log servers $N$ that must issue an SCT for a certificate to be considered valid and the number of trusted map servers $M$ that the relying party requires to respond.
Hence, \textbf{an attack on the \geocert{} fetching objective}, i.e., not returning the complete set of certificates, \textbf{is prevented if at most $\bm{N-1}$ log servers and at most $\bm{M-1}$ map servers are malicious}.

Finally, once the relying party has received the set of all \geocerts{} for a given space or identifier, it determines, based on this set, whether a received cryptographic object of interest, e.g., a TLS certificate, is valid or not.
\newtheorem*{objectivefilter}{Objective (Validation)}
\greyborder{
  \begin{objectivefilter}\label{objective:filter}
    A relying party can, based on its trust preference and the set of all \geocerts{} in a certain space or with a certain identifier, deterministically restrict the set of valid cryptographic objects.
  \end{objectivefilter}
}
Based on its trust preference, the relying party selects the relevant \geocerts{}, i.e., the \geocerts{} that are signed by the geo CAs with the highest trust level.
This ensures that all \geocerts{} issued by a malicious geo CA are ignored, as long as there exists a \geocert{} issued by a more highly trusted geo CA\@.
The cryptographic object of interest is compared to the attributes contained in these \geocerts{}, i.e., if the object conflicts with any attribute in any \geocert{}, it is rejected.

\myparagraph{Split-View Attacks}
In split-view attacks, a trusted log server presents the map server (and thus indirectly the relying party) with a (correctly) signed tree root of a \emph{different} Merkle tree to hide \geocerts{}.
While this attack cannot be prevented, it is very difficult to perform stealthily and can easily be detected if map servers (and geo CAs) gossip recent STHs among themselves.
Similarly, a map server performing a split-view attack and presenting different views to relying parties is prevented by the append-only consistency tree, which never removes an SMH.
A map server doing a split-view attack must continue to maintain multiple consistency trees and ensure that relying parties and auditors of different versions never compare their trees.
Hence, standard solutions like CT monitors or gossiping can help detect split view attacks~\cite{Chuat2015,Dahlberg2018a,Oxford2020}.

\section{Location Verification}\label{sec:location-verification}
The location verification system needs to be robust, preventing criminals from registering arbitrarily many unclaimed locations to give weight to spurious certificates.

We analyze the suggestions made by Kim et al.~\cite{Kim2012} and elaborate on why we deem them unsatisfactory. They propose a hierarchical verification system in which certificates of higher verification levels are preferred over lower levels. The four proposed levels in descending weights are: (1) CA-Extended Validation-signed, requiring a CA employee to travel to the physical space and verify the association; (2) CA-Location Validation-signed, where the registering entity is required to travel to a CA and prove the association through documents (e.g., a water bill or lease) or by retrieving a code sent by physical mail; (3) CA-signed, where no ownership proof is required; and (4) self-signed.

We deem in-person verification an appropriate but expensive mechanism. Officials or CA employees could personally verify the \emph{legitimacy of foreign actors}, \emph{land ownership claims} (also through the use of specialized registers), or \emph{the veracity of location statements}. Note that \fgpki{} supports all of these different types of claims by including the type in the \geocert{}. Once such verification has occurred, hierarchical delegation, as already done in today's web PKI, is a natural extension. A mall could for instance delegate its subspaces and provide the verification by proxy to the CAs which sign the GeoCerts. This would result in a clear declaration of which stores, WiFi hotspots, bank ATMs, etc. are expected in the space, hardening it against \emph{payment fraud} and \emph{fake services}.

We do not deem self-signing and weak verifications, such as challenges sent via physical mail, acceptable, as they are easily faked.
A legitimate occupant may overrule spurious \geocerts{} by registering a \geocert{} with a higher level of verification.
However, this only works as long as they can be made aware of these spurious claims, and there is an incentive for the occupant to claim the space, which is not always given.
The occupant may not have any website to claim the space and may not want to spend resources to displace the scammers without direct gains for themselves.
Nonetheless, low-effort certificate issuance similar to Let's Encrypt would be beneficial for fast adoption.
Hence, the question arises: \emph{what are the options for enabling low-effort verifications and how can we mitigate the resulting lower level of trust?}

Although an in-depth discussion and evaluation of low-effort verification mechanisms is beyond the scope of this work, we provide initial observations and directions, leaving an in-depth treatment to future research.
Regarding the problem of less reliable location verification mechanisms, \fgpki{} offers a deterministic and flexible solution to support mechanisms with different levels of trust.

The achievable security is typically \emph{proportional} to the level of trained human involvement in the verification process. A trained CA employee traveling to the location to verify legitimate occupancy is harder to deceive than a fully automated remote system.
Similarly, for the price, a fully automated system will be cheaper in the long term, while any human training and employment adds to the cost.

\subsection{Alternative Verifiers}

One may consider using existing infrastructure or pre-existing human movement, such as everyday commutes, to automate the process and reduce specialized training requirements.
For example, in countries with a comprehensive postal system, mail is delivered to every house.
An approach is to have postal employees perform location verifications that are on their postal route.
An alternative in densely populated areas may be a crowd sourcing approach, by which people could incidentally verify location claims on their commute, akin to the volunteer data gathering done by projects like Open Street Maps.
On the fully automated end of the spectrum, drones may be a viable option. Although still futuristic, there are already carrier drones for specialized applications (e.g., time-critical transportation of blood probes~\cite{drones6080195}), and could be a viable option as they become more common.

\subsection{Verification Medium}

We envision that verification may be carried out by CAs issuing challenges to occupants. They would in turn perform a challenge response by physically displaying the response at the location, which would then be checked by the verifier.
This requires the physically displayed challenge to be digitized.

We list two possibilities for this.
The first option would be \textbf{optically}, through the display of a \textbf{QR code} containing the challenge response.
An application that scans the QR code and automatically verifies the signature could be provided.
This takes advantage of commodity hardware in the form of cell phones and is an intuitive mechanism. 
The difficulty lies in finding a spot where the QR code can be displayed, which is easily accessible for the verifier, while simultaneously under the occupant's control, such as pasted to the inside of a closed window.
Note that it must be difficult for an attacker to set up or replace the displayed code.
Depending on the concrete location, it may be challenging to meet these requirements.
Additionally, finding and scanning a QR code requires additional work by the verifier.

The second option would be through a \textbf{wireless protocol}. This is more convenient for the verifier as it can be more highly automated and may even reach the level of passive sensing that applications in the smart city domain are already striving for today~\cite{smartcities4010020}.
The challenge lies in the exact positioning and detection of the beacon, through a \textbf{secure distance bounding} protocol~\cite{leu_message_2020}.
While they exist for specialized applications and are more common for short range communication, the implementation on commodity hardware and especially cell phones (typically through the use of Ultra Wide Band technology) is an active area of research producing a steady stream of both attacks~\cite{leu_ghost_2021, anliker_time_2023} and countermeasures~\cite{singh_uwb-ed_2019, singh_security_2021, v-range_2023}.
It seems likely that this option will be viable in the near future.

\subsection{Security Considerations}\label{sec:location-verification:security-considerations}

Low-effort location verification methods must be supported with caution due to their abuse potential by scammers.
However, supporting them is imperative for accessibility and to drive early adoption.
However, there is a tricky balance between adoptability (facilitated by simple early sign-up mechanisms) and trustworthiness: simple adoption may result in early scams, which erodes trustworthiness of the system.
With the flexible trust model of \fgpki{}, inspired by \fpkiorig{}, relying parties can set their own preferences for verification methods,
with their own strategies to handle the trade off.
Space owners likewise can steer which verification methods are acceptable through policies applied as attributes to their children, inherited by subspaces of that space.

This allows for a varying array of use cases, from the early tinkerer testing the system without much concern for security, to the client-side preference meant to protect users from ceding their personal information to a physical phishing scam.
In \fgpki{}, this can be solved as follows:
Each \geocert{} must contain the type of location verification that the CA performed before submitting it to the system.
The client then specifies, in its trust preference, in addition to the CAs' trust levels, the type of validation, i.e., $\text{TP} = {\langle \text{CA}_i, \text{LocVer}_i, V_i, \text{TL}_i\rangle}_{i\in [1,N]}$.
This enables less secure location verification mechanisms to coexist with more secure mechanisms without impacting their security by preventing downgrade attacks.
Less secure verification mechanisms are only valid if the parent policy and the client's preference allow them.
We envision the emergence of preferences and policies recommended per use-case, akin to the current list of trusted CAs distributed by the browser vendors.
This information may also serve as a filter should there be a need to reduce overhead, i.e., geo map servers could opt to only support reliable location verification mechanisms.

\section{Related Work}
In this section, we review related work on other (geographical) PKIs and location privacy.

\subsection{Public Key Infrastructures}
\fgpki{} takes inspiration from \fpkiorig{}~\cite{Chuat2022}, in particular regarding the idea that the validation should be performed at the relying party based on a local and individual trust preference.
The main difference and challenges of this work are the manipulation of spatial data, i.e., finding sensible geographical projections, combining different space claim validation mechanisms, and ensuring the completeness of returned \geocerts{} while providing efficient spatial indexing.
Although our focus is on the web PKI, the \fgpki{} framework is flexible and extensible and could enhance other PKIs with geographical information, such as RPKI~\cite{RFC8210} or DNSSEC~\cite{RFC4033}.

\subsection{Location Privacy}
Location data, especially time series, is considered sensitive user information~\cite{Golle2009,Gambs2010,Freudiger2012,DeMontjoye2013,Backes2017}.
Since the map server learns the location of the relying parties through queries, this may raise privacy concerns.
The sporadic nature of \fgpki{} queries and the fact that many users will perform queries for the same locations, i.e., in places where \geocerts{} are commonly used, limit the inferable information.
Increasing the query radius, querying using the coarse grid defined by the SMT, or relaying the queries via proxies can offer additional protection.
There is extensive research on mechanisms to preserve privacy for location-based services~\cite{Andres2013,Bordenabe2014,Shokri2015,Jiang2021}.
\fgpki{} clients could be enhanced with similar mechanisms to preserve user privacy.

\subsection{GeoPKI}
Kim et al.~\cite{Kim2012} propose a geographical PKI (\geopkiorig{}) based on special certificates containing geographical information, called \emph{GeoCerts}.
The \geopkiorig{} database then provides a mapping from the geographic coordinates to the GeoCert valid at this position.
Depending on the effort of the certificate issuance process, GeoCerts are assigned different security levels, e.g., preferring extended validation (EV) certificates to self-signed certificates.
\geopkiorig{} considers the globe as a unit cube and allows separating it into arbitrarily small cubes by applying a 3D-tree, i.e., splitting along the three axes one by one.
Each of these cubes is then assigned a single valid GeoCert.
Furthermore, they make use of hierarchically structured MHTs to delegate space ownership.

Unfortunately, their system has several limitations in terms of practicality and usability.
It does not allow the presence of multiple GeoCerts (issued by different CAs) at one location.
Their Cartesian coordinate system leads to inefficiencies by covering large spaces that are unlikely to require certificates, e.g., the Earth's core, and by not aligning cubes to the Earth's surface, increasing the number of nodes required to represent common spaces, e.g., a house or shop.
Additionally, by directly tying physical space to public keys, inconsistencies with other PKIs may occur.
Finally, revocation is done by claiming the space with a GeoCert of a higher security level, which means that once an EV GeoCert, which has the highest security level, exists, it cannot be revoked by this CA anymore (unless signatures from \emph{multiple} CAs overwrite the GeoCert).

We innovate on the basic concept of a PKI based on a geographically structured MHT and address these deficiencies.
\fgpki{}'s data structure provides significant improvements in terms of efficiency and real-world feasibility, and we provide a prototype implementation and analyze its deployability, in contrast to the purely theoretical work of Kim et al.

In particular, \fgpki{} minimizes the number of MHT nodes (and their tree depth) associated with a certificate or query, reducing processing, storage, and network overhead.
In \fgpki{}, multiple certificates can be assigned to a single node, which (1) avoids the need for precise borders between neighboring certificates consisting of many nodes (instead, fewer border nodes containing both certificates are used), and (2) keeps \fgpki{} future-proof, i.e., supports novel use cases that require new overlapping certificates or more fine-grained space claims.
While GeoPKI mandates cube-like shapes, \fgpki{} supports frustums with variable heights (including cube-like shapes) aligned to the earth's surface, which minimizes the number of nodes for common shapes such as houses and shops as well as large areas such as airports and countries.

Regarding the tree depth of nodes, the covered space is reduced by a factor of roughly \num{64} compared to GeoPKI, reducing MHT proof lengths by roughly \num{6} fold, and \fgpki{} supports efficient space claims for entities claiming an area at all altitudes and relying parties without accurate altitude information.

\section{Conclusion}

With the increasing digitalization of real-world processes, the relevance of the web PKI continues to grow, as it establishes trust in diverse use cases.
Recognizing that location information is intuitive and valuable for users to establish trust, we explore the technical challenges of integrating it into the web PKI.
\fgpki{} demonstrates that the web PKI can be enriched with location information in a scalable manner today and only requires standardization of \geocerts{} and deployment of a geo map server.
Providing location information through \fgpki{} addresses use cases with no prior solutions since location information is fundamentally different from the domain name information currently provided by the web PKI.
Once deployed, we anticipate a plethora of new use cases to emerge, offering enhanced security through location information.

\section*{Acknowledgments}
We gratefully acknowledge support from Princeton University, ETH Zurich, the Zurich Information Security and Privacy Center (ZISC), and the Werner Siemens-Stiftung (WSS) Centre for Cyber Trust at ETH Zurich.

\printbibliography

@inproceedings{BirgeLee2018,
  author = {Henry Birge-Lee and Yixin Sun and Anne Edmundson and Jennifer Rexford and Prateek Mittal},
  booktitle = {Proceedings of the USENIX Security Symposium},
  title = {Bamboozling Certificate Authorities with {BGP}},
  year = {2018},
  month = aug,
  isbn = {978-1-939133-04-5},
}

@InCollection{Kim2012,
  author       = {Tiffany Hyun-Jin Kim and Virgil Gligor and Adrian Perrig},
  booktitle    = {European Workshop on Public Key Infrastructures, Services and Applications (EuroPKI)},
  title        = {{GeoPKI}: Converting Spatial Trust into Certificate Trust},
  year         = {2012},
  doi          = {10.1007/978-3-642-40012-4_9},
}

@InProceedings{Chuat2022,
  author       = {Laurent Chuat and Cyrill Krähenbühl and Prateek Mittal and Adrian Perrig},
  booktitle    = {Proceedings of the Network and Distributed System Security Symposium (NDSS)},
  title        = {F-{PKI}: Enabling Innovation and Trust Flexibility in the {HTTPS} Public-Key Infrastructure},
  year         = {2022},
  doi          = {10.14722/ndss.2022.24241},
}

@inproceedings{Dai2021,
  author       = {Tianxiang Dai and Haya Shulman and Michael Waidner},
  booktitle    = {Proceedings of the ACM Conference on Computer and Communications Security (CCS)},
  title        = {Let's Downgrade {Let's} {Encrypt}},
  year         = {2021},
  month        = nov,
  doi          = {10.1145/3460120.3484815},
}

@techreport{Hoogstraaten2012,
  author       = {Hans Hoogstraaten and Ronald Prins and Daniël Niggebrugge and Danny Heppener and Frank Groenewegen and Janna Wettink and Kevin Strooy and Pascal Arends and Paul Pols and Robbert Kouprie and Steffen Moorrees and Xander van Pelt and Yun Zheng Hu},
  title        = {{Black Tulip}: Report of the investigation into the {DigiNotar} Certificate Authority breach},
  year         = {2012},
  month        = aug,
  doi          = {10.13140/2.1.2456.7364},
}

@Misc{Comodo2011,
  author       = {{Comodo}},
  url          = {https://perma.cc/AT8Q-TJJC},
  title        = {Report of Incident: {Comodo} detected and thwarted an intrusion on 26-MAR-2011},
  year         = {2011},
  month        = mar,
}

@misc{FIDO2022,
  author       = {{FIDO Alliance}},
  title        = {{Fast} {IDentity} {Online} ({FIDO}) Standard},
  year         = {2022},
  url          = {https://fidoalliance.org},
}

@misc{sslmate2022,
  author       = {{sslmate}},
  title        = {Timeline of Certificate Authority Failures},
  year         = {2022},
  month        = oct,
  url          = {https://perma.cc/VZ3E-GMW9},
}

@techreport{wgs84,
  author       = {{Office of Geomatics}},
  title        = {Department of Defense World Geodetic System 1984: Its Definition and Relationships with Local Geodetic Systems},
  institution  = {National Geospatial-Intelligence Agency ({NGA})},
  type         = {Standard},
  year         = {2014},
  month        = jul,
}

@online{UsDepartmentCommerceHighestPoint,
  title        = {What is the highest point on Earth as measured from Earth's center?},
  url          = {https://perma.cc/KUW8-42XS},
  author       = {{US Department of Commerce, National Oceanic and Atmospheric Administration}},
  date         = {2023-01-20},
  year         = {2023},
}

@article{Gardner2014,
  title        = {So, How Deep Is the {Mariana} Trench?},
  volume       = {37},
  doi          = {10.1080/01490419.2013.837849},
  number       = {1},
  journal      = {Marine Geodesy},
  author       = {Gardner, James V. and Armstrong, Andrew A. and Calder, Brian R. and Beaudoin, Jonathan},
  date         = {2014-01-02},
  year         = {2014},
}

@article{Kabat2005,
  title        = {Climate proofing the Netherlands},
  volume       = {438},
  doi          = {10.1038/438283a},
  number       = {7066},
  journal      = {Nature},
  author       = {Kabat, Pavel and van Vierssen, Wim and Veraart, Jeroen and Vellinga, Pier and Aerts, Jeroen},
  year         = {2005},
}

@techreport{Morton1966,
  title        = {A computer Oriented Geodetic Data Base; and a New Technique in File Sequencing},
  institution  = {{IBM} Ltd.},
  type         = {Technical Report},
  author       = {Morton, Guy Macdonald},
  year         = {1966},
}

@misc{Geohash,
  title        = {geohash.org is public!},
  url          = {https://perma.cc/8XZV-WXE5},
  author       = {Gustavo Niemeyer},
  date         = {2008-02-26},
  year         = {2008},
}

@misc{GeohashExplanation,
  title        = {Geohash Intro},
  url          = {https://perma.cc/U2U2-XUV5},
  author       = {Phil Whelan},
  year         = {2011},
  month        = dec,
}

@incollection{Brands1993,
  author       = {Stefan Brands and David Chaum},
  booktitle    = {Advances in Cryptology},
  publisher    = {Springer Berlin Heidelberg},
  title        = {Distance-Bounding Protocols},
  year         = {1993},
  doi          = {10.1007/3-540-48285-7_30},
}

@article{Yan2014,
  author       = {Shihao Yan and Robert Malaney and Ido Nevat and Gareth W. Peters},
  journal      = {{IEEE} Transactions on Vehicular Technology},
  title        = {Optimal Information-Theoretic Wireless Location Verification},
  year         = {2014},
  month        = sep,
  number       = {7},
  volume       = {63},
  doi          = {10.1109/tvt.2014.2302022},
}

@article{Vora2006,
  author       = {Adnan Vora and Mikhail Nesterenko},
  journal      = {{IEEE} Transactions on Dependable and Secure Computing},
  title        = {Secure Location Verification Using Radio Broadcast},
  year         = {2006},
  month        = oct,
  number       = {4},
  volume       = {3},
  doi          = {10.1109/tdsc.2006.57},
}

@inproceedings{Singelee2005,
  author       = {D. Singelee and B. Preneel},
  booktitle    = {{IEEE} International Conference on Mobile Adhoc and Sensor Systems Conference},
  title        = {Location verification using secure distance bounding protocols},
  year         = {2005},
  doi          = {10.1109/mahss.2005.1542879},
}

@article{Nosouhi2020,
  author       = {Mohammad Reza Nosouhi and Shui Yu and Wanlei Zhou and Marthie Grobler and Habiba Keshtiar},
  journal      = {Journal of Parallel and Distributed Computing},
  title        = {Blockchain for secure location verification},
  year         = {2020},
  month        = feb,
  volume       = {136},
  doi          = {10.1016/j.jpdc.2019.10.007},
}

@article{Capkun2008,
  author       = {Srdjan Capkun and Kasper Rasmussen and Mario Cagalj and Mani Srivastava},
  journal      = {{IEEE} Transactions on Mobile Computing},
  title        = {Secure Location Verification with Hidden and Mobile Base Stations},
  year         = {2008},
  month        = apr,
  number       = {4},
  volume       = {7},
  doi          = {10.1109/tmc.2007.70782},
}

@inproceedings{Chuat2015,
  author       = {Laurent Chuat and Pawel Szalachowski and Adrian Perrig and Ben Laurie and Eran Messeri},
  booktitle    = {Proceedings of the {IEEE} Conference on Communications and Network Security ({CNS})},
  title        = {Efficient gossip protocols for verifying the consistency of Certificate logs},
  year         = {2015},
  month        = sep,
  doi          = {10.1109/cns.2015.7346853},
}

@incollection{Dahlberg2018a,
  author       = {Rasmus Dahlberg and Tobias Pulls},
  booktitle    = {Secure {IT} Systems},
  publisher    = {Springer International Publishing},
  title        = {Verifiable Light-Weight Monitoring for Certificate Transparency Logs},
  year         = {2018},
  doi          = {10.1007/978-3-030-03638-6_11},
}

@inproceedings{Oxford2020,
  author       = {Michael Oxford and David Parker and Mark Ryan},
  booktitle    = {Proceedings of the {IEEE} Conference on Communications and Network Security ({CNS})},
  title        = {Quantitative Verification of Certificate Transparency Gossip Protocols},
  year         = {2020},
  month        = jun,
  doi          = {10.1109/cns48642.2020.9162197},
}

@Article{drones6080195,
  author       = {Niglio, Fabrizio and Comite, Paola and Cannas, Andrea and Pirri, Angela and Tortora, Giuseppe},
  title        = {Preliminary Clinical Validation of a Drone-Based Delivery System in Urban Scenarios Using a Smart Capsule for Blood},
  journal      = {Drones},
  volume       = {6},
  year         = {2022},
  number       = {8},
  doi          = {10.3390/drones6080195},
}

@Article{smartcities4010020,
  author       = {Dooley, Ken},
  title        = {Direct Passive Participation: Aiming for Accuracy and Citizen Safety in the Era of Big Data and the Smart City},
  journal      = {Smart Cities},
  volume       = {4},
  year         = {2021},
  number       = {1},
  doi          = {10.3390/smartcities4010020}
}

@inproceedings{Backes2017,
  author       = {Backes, Michael and Humbert, Mathias and Pang, Jun and Zhang, Yang},
  title        = {walk2friends: Inferring Social Links from Mobility Profiles},
  year         = {2017},
  doi          = {10.1145/3133956.3133972},
  booktitle    = {Proceedings of the ACM Conference on Computer and Communications Security (CCS)},
  numpages     = {15},
}

@article{DeMontjoye2013,
  title        = {Unique in the Crowd: The privacy bounds of human mobility},
  volume       = {3},
  rights       = {2013 The Author(s)},
  doi          = {10.1038/srep01376},
  shorttitle   = {Unique in the Crowd},
  number       = {1},
  journal      = {Scientific Reports},
  shortjournal = {Sci Rep},
  author       = {de Montjoye, Yves-Alexandre and Hidalgo, César A. and Verleysen, Michel and Blondel, Vincent D.},
  date         = {2013-03-25},
  year         = {2013},
  langid       = {english},
}

@incollection{Golle2009,
  title        = {On the Anonymity of Home/Work Location Pairs},
  volume       = {5538},
  booktitle    = {Pervasive Computing},
  publisher    = {Springer Berlin Heidelberg},
  author       = {Golle, Philippe and Partridge, Kurt},
  year         = {2009},
  langid       = {english},
  doi          = {10.1007/978-3-642-01516-8_26},
}

@InProceedings{Freudiger2012,
  author       = {Freudiger, Julien and Shokri, Reza and Hubaux, Jean-Pierre},
  title        = {Evaluating the Privacy Risk of Location-Based Services},
  booktitle    = {Financial Cryptography and Data Security},
  year         = {2012},
  isbn         = {978-3-642-27576-0}
}

@inproceedings{Gambs2010,
  author       = {Gambs, Sébastien and Killijian, Marc-Olivier and del Prado Cortez, Miguel Núñez},
  title        = {Show Me How You Move and {I} Will Tell You Who You Are},
  year         = {2010},
  doi          = {10.1145/1868470.1868479},
  booktitle    = {Proceedings of the ACM SIGSPATIAL International Workshop on Security and Privacy in GIS and LBS},
  numpages     = {8},
}

@inproceedings{Andres2013,
  author       = {Andr\'{e}s, Miguel E. and Bordenabe, Nicol\'{a}s E. and Chatzikokolakis, Konstantinos and Palamidessi, Catuscia},
  title        = {Geo-Indistinguishability: Differential Privacy for Location-Based Systems},
  year         = {2013},
  doi          = {10.1145/2508859.2516735},
  booktitle    = {Proceedings of the ACM Conference on Computer and Communications Security (CCS)},
  numpages     = {14},
}

@inproceedings{Bordenabe2014,
  author       = {Bordenabe, Nicol\'{a}s E. and Chatzikokolakis, Konstantinos and Palamidessi, Catuscia},
  title        = {Optimal Geo-Indistinguishable Mechanisms for Location Privacy},
  year         = {2014},
  doi          = {10.1145/2660267.2660345},
  booktitle    = {Proceedings of the ACM Conference on Computer and Communications Security (CCS)},
  numpages     = {12},
}

@article{Shokri2015,
  title        = {Privacy Games: Optimal User-Centric Data Obfuscation},
  volume       = {2015},
  doi          = {10.1515/popets-2015-0024},
  shorttitle   = {Privacy Games},
  number       = {2},
  journal      = {Proceedings on Privacy Enhancing Technologies},
  author       = {Shokri, Reza},
  date         = {2015-06-01},
  year         = {2015},
  langid       = {english}
}

@article{Jiang2021,
  author       = {Jiang, Hongbo and Li, Jie and Zhao, Ping and Zeng, Fanzi and Xiao, Zhu and Iyengar, Arun},
  title        = {Location Privacy-Preserving Mechanisms in Location-Based Services: A Comprehensive Survey},
  year         = {2021},
  issue_date   = {January 2022},
  volume       = {54},
  number       = {1},
  doi          = {10.1145/3423165},
  journal      = {ACM Comput. Surv.},
  month        = jan,
  articleno    = {4},
  numpages     = {36}
}

@inproceedings{Song2010a,
  author       = {Yimin Song and Chao Yang and Guofei Gu},
  booktitle    = {Proceedings of the IEEE/IFIP International Conference on Dependable Systems and Networks (DSN)},
  title        = {Who is peeping at your passwords at Starbucks? To catch an evil twin access point},
  year         = {2010},
  month        = jun,
  doi          = {10.1109/dsn.2010.5544302},
}

@misc{CIAPakistan2021,
  author       = {Lori Uildriks},
  title        = {How a covert {CIA} operation led to vaccine hesitancy in Pakistan},
  url          = {https://perma.cc/ZR4X-C7UJ},
  year         = {2021},
  month        = may,
}

@article{Mukherjee2021,
  author       = {Dattatreya Mukherjee and Upasana Maskey and Angela Ishak and Zouina Sarfraz and Azza Sarfraz and Vikash Jaiswal},
  journal      = {Postgraduate Medical Journal},
  title        = {Fake {COVID}-19 vaccination in {India}: an emerging dilemma?},
  year         = {2021},
  month        = aug,
  number       = {e2},
  volume       = {98},
  doi          = {10.1136/postgradmedj-2021-141003},
}

@inproceedings{leu_message_2020,
  title        = {Message {Time} of {Arrival} {Codes}: {A} {Fundamental} {Primitive} for {Secure} {Distance} {Measurement}},
  shorttitle   = {Message {Time} of {Arrival} {Codes}},
  doi          = {10.1109/SP40000.2020.00010},
  booktitle    = {Proceedings of the IEEE Symposium on Security and Privacy (S\&P)},
  author       = {Leu, Patrick and Singh, Mridula and Roeschlin, Marc and Paterson, Kenneth G. and Čapkun, Srdjan},
  month        = may,
  year         = {2020},
}

@inproceedings{singh_security_2021,
  title        = {Security analysis of {IEEE} 802.15.4z/{HRP} {UWB} time-of-flight distance measurement},
  doi          = {10.1145/3448300.3467831},
  booktitle    = {Proceedings of the {ACM} {Conference} on {Security} and {Privacy} in {Wireless} and {Mobile} {Networks} (WiSec)},
  author       = {Singh, Mridula and Roeschlin, Marc and Zalzala, Ezzat and Leu, Patrick and Čapkun, Srdjan},
  month        = jun,
  year         = {2021},
}

@inproceedings{singh_uwb-ed_2019,
  title        = {{UWB}-{ED}: Distance Enlargement Attack Detection in Ultra-Wideband},
  isbn         = {978-1-939133-06-9},
  booktitle    = {Proceedings of the USENIX Security Symposium},
  author       = {Singh, Mridula and Leu, Patrick and Abdou, AbdelRahman and Čapkun, Srdjan},
  year         = {2019},
}

@misc{leu_ghost_2021,
  title        = {Ghost {Peak}: {Practical} {Distance} {Reduction} {Attacks} {Against} {HRP} {UWB} {Ranging}},
  shorttitle   = {Ghost {Peak}},
  doi          = {10.48550/arXiv.2111.05313},
  publisher    = {arXiv},
  author       = {Leu, Patrick and Camurati, Giovanni and Heinrich, Alexander and Roeschlin, Marc and Anliker, Claudio and Hollick, Matthias and Čapkun, Srdjan and Classen, Jiska},
  month        = nov,
  year         = {2021},
  note         = {arXiv:2111.05313 [cs]},
}

@misc{anliker_time_2023,
  title        = {Time for {Change}: {How} {Clocks} {Break} {UWB} {Secure} {Ranging}},
  shorttitle   = {Time for {Change}},
  doi          = {10.48550/arXiv.2305.09433},
  publisher    = {arXiv},
  author       = {Anliker, Claudio and Camurati, Giovanni and Čapkun, Srdjan},
  month        = may,
  year         = {2023},
  note         = {arXiv:2305.09433 [cs]},
}

@InProceedings{v-range_2023,
  author       = {Mridula Singh and Marc Roeschlin and Aanjhan Ranganathan and Srdjan Capkun},
  booktitle    = {Proceedings of the Symposium on Network and Distributed Systems Security (NDSS)},
  title        = {{V-Range}: Enabling Secure Ranging in {5G} Wireless Networks},
  year         = {2022},
  month        = apr,
  doi          = {10.14722/ndss.2022.23151},
}

@article{Schiavina2022,
  title        = {{GHS}-{POP} {R2022A} - {GHS} population grid multitemporal (1975-2030)},
  doi          = {10.2905/D6D86A90-4351-4508-99C1-CB074B022C4A},
  journal      = {European Commission, Joint Research Centre ({JRC})},
  author       = {Schiavina, Marcello and Freire, Sergio and {MacManus}, Kytt},
  date         = {2022-06-25},
  year         = {2022},
  langid       = {english},
}

@book{Freire2016,
  title        = {Development of new open and free multi-temporal global population grids at 250 m resolution},
  isbn         = {978-90-816960-6-7},
  publisher    = {Association of Geographic Information Laboratories in Europe ({AGILE})},
  author       = {Carneiro Freire, Sergio Manuel and Macmanus, Kytt and Pesaresi, Martino and Doxsey-Whitfield, Erin and Mills, Jane},
  year         = {2016},
  month        = dec,
}

@misc{Nasa2019,
  title        = {{ASTER} Global Digital Elevation Model V003},
  doi          = {10.5067/ASTER/ASTGTM.003},
  publisher    = {{NASA} {EOSDIS} Land Processes {DAAC}},
  author       = {{NASA/METI/AIST/Japan Spacesystems And U.S./Japan ASTER Science Team}},
  year         = {2019},
}

@misc{gecko-source-code,
  title        = {{GECKO} Source Code and Evaluation Results},
  author       = {Nico Hauser},
  url          = {https://github.com/netsec-ethz/geopki},
  year         = 2022,
}

@techreport{RFC4033,
  title = {{DNS Security Introduction and Requirements}},
  author = {Arends, R. and Austein, R. and Larson, M. and Massey, D. and Rose, S.},
  type = {RFC},
  number = {4033},
  institution = {IETF},
  month = mar,
  year = 2005,
  doi = {10.17487/RFC4033},
}

@techreport{RFC6749,
  title = {{The OAuth 2.0 Authorization Framework}},
  author = {Hardt, D.},
  type = {RFC},
  number = {6749},
  institution = {IETF},
  month = oct,
  year = 2012,
  doi = {10.17487/RFC6749},
}

@techreport{RFC6962,
  title = {{Certificate Transparency}},
  author = {Laurie, B. and Langley, A. and Kasper, E.},
  type = {RFC},
  number = {6962},
  institution = {IETF},
  month = jun,
  year = 2013,
  doi = {10.17487/RFC6962},
}

@techreport{RFC8210,
  title = {{The Resource Public Key Infrastructure (RPKI) to Router Protocol, Version 1}},
  author = {Bush, R. and Austein, R.},
  type = {RFC},
  number = {8210},
  institution = {IETF},
  month = sep,
  year = 2017,
  doi = {10.17487/RFC8210},
}

\begin{appendices}
  \crefalias{section}{appendix}
  \section{SMT Formalized}\label{app:smt-formalized}

This section formally describes the SMT from \cref{sec:smt}.
First, we fix the earth model and then derive the discretization precision.

\subsection{Earth Model}\label{app:smt-formalized:earth-model}

We use the earth ellipsoid model of the world geodetic system from 1984 (WGS84)~\cite{wgs84}.
WGS84 defines the \emph{semi-major axis} (half of the largest diameter of an ellipsoid, at the equator) $r_a$ to be \SI{6378137.0}{\meter}. The \emph{semi-minor axis} (the polar radius) $r_b$ is defined as \SI{6356752.3142}{\meter}.
For referencing points on the planet's two-dimensional surface, WGS84 defines the (standard) geodetic coordinates \emph{longitude} and \emph{latitude} with values in the range $[-180, 180]$ and $[-90, 90]$, respectively.
For the altitude, WGS84 uses the EGM2008 datum.

Each of the three coordinates is discretized in a way that preserves meter-precision in the worst case.
With the semi-major axis $r_a$, the maximum circumference of the ellipsoid is $2 r_a \pi$~meters.
Using base two, $B_{x}:= \left\lfloor \log_2(2 \cdot r_a \cdot \pi) \right\rfloor + 1 = 26$ bits thus suffice to represent the longitude to a meter precision.
Let $C_{x} := 2^{B_{x}} - 1$ denote the maximum discretized value for the longitude.
The circumference along the semi-minor axis is $2 r_b \pi$~meters, but we only need to represent half of it, i.e. $r_b \pi$, analogous to the latitude value range.
Therefore $\left\lfloor\log_2(r_b \cdot \pi)\right\rfloor + 1 = 25 =: B_{y}$ bits suffice to represent each possible meter value for the latitude.
Analogously to $C_{x}$, let $C_y := 2^{B_{y}} - 1$ be the maximum discretized value for the latitude.

For the altitude, we fix bounds due to the absence of a natural upper bound and the assumption that there is limited value in the ability to represent coordinates within the earth.
At the time of writing, the highest point on earth is Mount Everest reaching an altitude just below \SI{9000}{\meter} above sea level~\cite{UsDepartmentCommerceHighestPoint}.
We assume that the Mariana Trench, at about \SI{11000}{\meter} below sea level~\cite{Gardner2014}, is a bound not exceeded in the near future.
The necessity of allowing volumes below sea level arises from countries such as the Netherlands where parts of the surface's altitude is below sea level~\cite{Kabat2005}.
Therefore we assume the altitude range starting at a depth of $D:=$\SI{-11000}{\meter} up to an altitude of \SI{9000}{\meter} above sea level. This should cover any current and near-future needs.
Covering this range requires $B_{z} := \left\lfloor \log_2(\num{9000} - D) \right\rfloor + 1 = 15$ bits.
More precisely, using $B_z$ bits starting at $D$, will cover the range from $D$ to $D + (2^{B_{z}} - 1) =$ \num{21767}m $=: H$.
Since EGM2008's zero point is at mean sea level~\cite{wgs84}, no shift of the range $[D, H]$ is necessary.
Finally, let $C_z := H - D = 2^{B_{z}} - 1$, analogous to $C_x$ and $C_y$.

\subsection{SMT Nodes}\label{app:smt-formalized:nodes}

The set of all nodes, $\mathcal{N}$, is defined as follows:
\begin{equation}
	\mathcal{N} := \bigcup_{\substack{i \in \{0, 1, ..., B_x + B_y]\\j \in \{0, 1, ..., B_z\}}} \{0, 1\}^{i} \times \{0, 1\}^{j}
\end{equation}

where $\mathcal{N}_r := \left(\epsilon, \epsilon \right)$ is the root of the tree with $\epsilon$ representing the empty bit string.
The first element in the pair is the surface bit string whereas the second element is the altitude bit string.

Let $\mathcal{C}: \mathcal{N} \mapsto \mathcal{P}\left(\mathcal{N}\right)$ be the function mapping each node to its children in the SMT where $\mathcal{P}$ denotes the power set.
Moreover let $||$ denote the concatenation operator.
For some node $n = (n_{xy}, n_{z})$, $\mathcal{C}(n)$ is defined as follows:

For $n_{z} = \epsilon$ :
\begin{align*}
	\mathcal{C}\left(\left(n_{xy}, \epsilon\right)\right) :=\:& \left\{(n_{xy} || 0, \epsilon), (n_{xy} || 1, \epsilon)\right\}\\
	\cup & \left\{(n_{xy}, 0), (n_{xy}, 1) \right\}
\end{align*}

For $n_{z} \in \bigcup_{j \in \{1,2,...,B_z-1\}} \{0,1\}^{j}$ :
\begin{align*}
	\mathcal{C}\left(\left(n_{xy}, n_{z}\right)\right) :=\:& \left\{(n_{xy}, n_{z} || 0), (n_{xy}, n_{z} || 1) \right\}
\end{align*}

And for leaves ($n_{z} \in \{0,1\}^{B_z}$)
\begin{align*}
	C\left(\left(n_{xy}, n_{z}\right)\right) := \{ \}
\end{align*}

\subsubsection{Hashes}\label{app:smt-formalized:hashes}

The tree dividing the earth's volume is extended to a Sparse Merkle Hash Tree (SMT) by defining the function $\mathcal{H}: \{0,1\}^{*} \times \{0,1\}^{*} \mapsto \{0,1\}^{256}$ mapping each possible bit string pair to a hash value.
For bit string pairs not corresponding to a node in the tree, $\mathcal{H}$ is defined to return a default value, i.e.,
\begin{align}
	\forall n \in \left(\{0,1\}^{*} \times \{0,1\}^{*} \right) \backslash \: \mathcal{N}. \: \mathcal{H}(n) := H(\mathbbzero)
\end{align}
where $\mathbbzero$ denotes the zero byte and $H: \{0,1\}^{*} \mapsto \{0,1\}^{256}$ is the SHA-256 hash function.

The data stored at nodes is a sequence of certificates denoted by $\mathcal{D}: \mathcal{N} \mapsto \left(c_i\right)_{i \in \mathbb{N}}$ where $\mathcal{L}: \mathcal{N} \mapsto \mathbb{N}$ for each node denotes the number of certificates in this sequence.
The sequence of certificates is sorted lexicographically by their SHA-256 hashes in ascending order.

As previously mentioned, this means that certificates are not exclusively associated with the tree's leaves but can instead be associated with any node in the tree.

A node $l$ is called a \emph{leaf} if it does not have any children in the SMT ($\mathcal{C}(l)=\emptyset$), i.e., it is a node whose $z$ bit string is of length $B_z$ (see \cref{sec:smt}).
The hash of a leaf is defined as follows:

$\forall l \in \{v \in \mathcal{N} \: | \: \mathcal{C}(v) = \emptyset \}$
\begin{align}
\label{eq:hash-leaf}
\begin{split}
	\mathcal{H}(l) &:= H\left(\mathbbzero \: || \: \concat_{i \in \mathcal{L}(l)} H\left(\mathcal{D}(l)_i\right) \right)\\
	\\% 
	&= H\left(\mathbbzero \:||\: H(c_0) \:||\: H(c_1) \:||\: ... \:||\: H(c_{\mathcal{L}(n)-1}) \right)
\end{split}
\end{align}

Note that for $\mathcal{L}(l) = 0$, this formula results in the hash value $H(\mathbbzero)$, the same value assigned to bit string pairs that are not part of the SMT.

The hash value of intermediate nodes depends on whether they are \emph{sparse}.
We define the predicate $\sigma \: : \: \mathcal{N} \mapsto \{\bot, \top\}$ for each node in the SMT to be $\top$ if and only if the respective node is sparse.
The definition is recursive; a leaf is sparse if it has no associated certificates, and an intermediate node is sparse if and only if it has no associated certificates and all its children are sparse (i.e., the whole subtree is empty).

For a leaf, the predicate is defined as follows:

$\forall l \in \{v \in \mathcal{N} \: | \: \mathcal{C}(v) = \emptyset \}$
\begin{align}
\label{eq:sparse-predicate:leafs}
\sigma(l) &= \left\{
\begin{array}{ll}
	\top & \text{if } \mathcal{L}(l) = 0 \\
	\bot & \text{otherwise}
\end{array}
\right.
\end{align}

Analogously, for intermediate nodes (non-leaf nodes):

$\forall n \in \{v \in \mathcal{N} \: | \: \mathcal{C}(v) \neq \emptyset \}$
\begin{align}
\label{eq:sparse-predicate:intermediate}
\sigma(n) &= \left\{
\begin{array}{ll}
	\top & \text{if } \mathcal{L}(n) = 0 \land \bigwedge\limits_{v \: \in \: \mathcal{C}(n)} \sigma(v) \\
	\bot & \text{otherwise}
\end{array}
\right.
\end{align}

For all sparse nodes, the hash value is defined to be the default value $H(\mathbbzero)$:

\begin{align}
	\forall v \in \{v \in \mathcal{N} \: | \: \sigma(v) \}. \: \mathcal{H}(v) := H(\mathbbzero)
\end{align}

This definition is consistent with the definition of hash values for leaves in \cref{eq:hash-leaf}.
Next, the hash value for non-sparse intermediate nodes is defined.

To build a Merkle hash tree, for any non-sparse intermediate node, the hash function incorporates the hashes of the node's children.
For any non-sparse intermediate node $n$, let $h_{n,1}, h_{n,2}, h_{n,3}$, and $h_{n,4}$ be the hash values of $n$'s children as follows:

$\forall n \in \{v \in \mathcal{N} \: | \: \mathcal{C}(v) \neq \emptyset \land \neg \sigma(v) \}$
\begin{align}
\begin{split}
	h_{n,1} &:= \mathcal{H}\left(\left(n_{xy} \:||\: 0, n_{z}\right)\right)
	\\
	h_{n,2} &:= \mathcal{H}\left(\left(n_{xy} \:||\: 1, n_{z}\right)\right)
	\\
	h_{n,3} &:= \mathcal{H}\left(\left(n_{xy}, n_{z} \:||\: 0\right)\right)
	\\
	h_{n,4} &:= \mathcal{H}\left(\left(n_{xy}, n_{z} \:||\: 1\right)\right)
\end{split}
\end{align}

For non-sparse intermediate nodes that have a non-empty set of associated certificates, $h_{n,5}$ is defined as the hash of the concatenation of the sequence of certificate hashes.

$\forall n \in \{v \in \mathcal{N} \: | \: \mathcal{C}(v) \neq \emptyset \land \neg \sigma(v) \land \mathcal{L}(v) \neq 0 \}$
\begin{align}
	\label{eq:hash-certs-intermediate-node}
	h_{n,5} := H\left(\concat_{i \in \mathcal{L}(n)} H\left(\mathcal{D}(n)_i\right)\right)
\end{align}

The hash value of non-sparse intermediate nodes is then defined as follows:

$\forall n \in \{v \in \mathcal{N} \: | \: \mathcal{C}(v) \neq \emptyset \land \neg \sigma(v) \}$
\begin{align}
	\label{eq:hash-intermediate-node}
	\mathcal{H}(n) := \left\{
	\begin{array}{lr}
		H(\mathbbone || h_{n,1} || h_{n,2} || h_{n,3} || h_{n,4}) & \text{if $\mathcal{L}(n) = 0$}\\
		H(\mathbbone || h_{n,1} || h_{n,2} || h_{n,3} || h_{n,4} || h_{n,5}) & \text{if $\mathcal{L}(n) \neq 0$}
	\end{array}\right.
\end{align}
The constant $\mathbbone$ is the byte encoding the integer one, i.e., the byte in which only the least significant bit is set.

The constants $\mathbbzero$ and $\mathbbone$ are prefixed in the hash computation to prevent collisions, where a given hash $H(a || b || c || d)$ for some $a$, $b$, $c$, and $d$ could simultaneously refer to a leaf with certificate hashes $a || b || c || d$ or an intermediate node with children hash values $a$, $b$, $c$, and $d$.

\subsubsection{Mapping a Spatial Point to a Bit String Pair}\label{app:smt-formalized:mapping-spatial-point}
\mbox{}\\ % linebreak, no lonely 'A' on a line
A point tuple $(x,y,z)$ at longitude $x$, latitude $y$, and altitude $z$ with
\begin{align*}
	(x, y, z) &\in [-180, 180] \times [-90, 90] \times [D, H+1]
\end{align*}

is first discretized to integer coordinates $(\hat{x},\hat{y}, \hat{z})$ as follows:

\begin{align}
\hat{x} := &
\left\{
\begin{array}{ll}
\left\lfloor
\frac{x + 180}{360} \cdot 2^{B_x}
\right\rfloor
& \text{if $x \in [-180, 180)$}
\\[1ex]
C_x & \text{if $x=180$}
\end{array}
\right.
\end{align}

\begin{align}
\hat{y} := &
\left\{
\begin{array}{ll}
\left\lfloor
\frac{y + 90}{180} \cdot 2^{B_y}
\right\rfloor
& \text{if $y \in [-90, 90)$}
\\[1ex]
C_y & \text{if $y = 90$}
\end{array}
\right.
\end{align}

\begin{align}
\hat{z} := &
\left\{
\begin{array}{ll}
\left\lfloor
z - D
\right\rfloor
& \text{if $z \in [D, H+1)$}
\\[1ex]
C_z & \text{if $z=H+1$}
\end{array}
\right.
\end{align}

with

\begin{align*}
	(\hat{x}, \hat{y}, \hat{z}) \in \left\{0, 1, ..., C_x\right\} \times
 \left\{0, 1, ..., C_y\right\} \times
 \left\{0, 1, ..., C_z\right\}
\end{align*}

The first factor in the longitude and latitude formulas is strictly smaller than $1$.
Moreover, $2^{B_x} = C_x + 1$ and $2^{B_y} = C_y + 1$ are one more than the maximum values that can be represented using $B_x$ and $B_y$ bits, respectively.
Therefore, the values resulting from the floor functions can always be represented using $B_x$ and $B_y$ bits.

The statement also holds for the altitude because subtracting $D$ changes the range from $[D,H+1)$ to $[0, H + 1 - D) = [0, C_z + 1)$, which, after rounding down, can by definition be represented using $B_z$ bits.

Let $(\tilde{x}, \tilde{y}, \tilde{z})$ be the big-endian integer binary encoding of $(\hat{x},\hat{y}, \hat{z})$ with
\begin{align*}
	(\tilde{x}, \tilde{y}, \tilde{z}) &\in \{0,1\}^{B_{x}} \times \{0,1\}^{B_{y}} \times \{0,1\}^{B_z}
\end{align*}

For any integers $i \in \{0, ..., B_x-1\}$, $j \in \{0, ..., B_y-1\}$ and $k \in \{0, ..., B_z-1\}$, let $\tilde{x}_{i}, \tilde{y}_{j}$ and $\tilde{z}_{k}$ denote the $i$-th, $j$-th and $k$-th bit of $\tilde{x}$, $\tilde{y}$ and $\tilde{z}$, respectively.
The first bit, i.e. $0$-th bit, is the most significant one.
The bit string pair corresponding of the point $p$ is then defined as
\begin{align}
	\label{eq:definition-bit-string-tuple}
	\left(\left(\concat_{i=0}^{B_{y} - 1} \tilde{x}_{i} || \tilde{y}_{i} \right) || \tilde{x}_{B_{x}-1}, \concat_{i=0}^{B_{z}-1} \tilde{z}_{i}\right)
\end{align}

\subsubsection{Geometric Interpretation of Nodes}\label{app:smt-formalized:geometric-interpretation}

For any node $n \in \mathcal{N}$, let $(n_{xy}, n_{z})$ be the corresponding bit string pair and let $n_{\tilde{x}}, n_{\tilde{y}}$ and $n_{\tilde{z}}$ be the big endian binary encoding of the three coordinates embedded in $n_{xy}$ and $n_{z}$.
The bit strings $n_{\tilde{x}}, n_{\tilde{y}}$ and $n_{\tilde{z}}$ correspond to the longitude, latitude and altitude, respectively.
The former two, $n_{\tilde{x}}$ and $n_{\tilde{y}}$, are obtained by de-interleaving $n_{xy}$ starting with an $x$ bit.
The bit string $\tilde{z}$ is equal to $n_{z}$.

Next let $l_{\tilde{x}}$, $l_{\tilde{y}}$ and $l_{\tilde{z}}$ be the lengths of $n_{\tilde{x}}$, $n_{\tilde{y}}$ and $n_{\tilde{z}}$, respectively.
Moreover let $n_{\tilde{x},i}$, $n_{\tilde{y},j}$ and $n_{\tilde{z},k}$ for any $i \in \{0, ..., l_{\tilde{x}}-1\}, j \in \{0, ..., l_{\tilde{y}}-1\}, k \in \{0, ..., l_{\tilde{z}}-1\}$ denote the $i$-th, $j$-th and $k$-th bit of $n_{\tilde{x}}, n_{\tilde{y}}$ and $n_{\tilde{z}}$, respectively.

\balance

Then $n_{\hat{x}}$, $n_{\hat{y}}$, $n_{\hat{z}}$ are the most-precise integers that can be re-constructed with the given bit string prefixes as follows:
\begin{align}
	\label{eq:intermediate-geometric-point-partial-height:x}
	n_{\hat{x}} :&= \sum_{i=0}^{l_{\tilde{x}} - 1} 2^{B_{x} - 1 - i} \cdot n_{\tilde{x}, i}\\
	\label{eq:intermediate-geometric-point-partial-height:y}
	n_{\hat{y}} :&= \sum_{i=0}^{l_{\tilde{y}} - 1} 2^{B_{y} - 1 - i} \cdot n_{\tilde{y}, i}\\
	\label{eq:intermediate-geometric-point-partial-height:z}
	n_{\hat{z}} :&= \sum_{i=0}^{l_{\tilde{z}} - 1} 2^{B_{z} - 1 - i} \cdot n_{\tilde{z}, i}
\end{align}

The corresponding \textbf{point} is given by:
\begin{align}
\begin{split}
\label{eq:intermediate-geometric-point-partial-height}
n_p :=\left(
n_{\hat{x}} \cdot \frac{360}{2^{B_x}} - 180,
n_{\hat{y}} \cdot \frac{180}{2^{B_y}} - 90,
D + n_{\hat{z}}
\right)
\end{split}
\end{align}

and the corresponding \textbf{volume} (a \emph{frustum}) by:

\begin{align}
\begin{split}
\label{eq:intermediate-geometric-volume-full-height}
n_V :=\left[
n_{\hat{x}} \cdot \frac{360}{2^{B_x}} - 180,
(n_{\hat{x}} + 2^{B_{x} - l_{\tilde{x}}}) \cdot \frac{360}{2^{B_x}} - 180
\right] &\times\\
\left[
n_{\hat{y}} \cdot \frac{180}{2^{B_y}} - 90,
(n_{\hat{y}} + 2^{B_{y} - l_{\tilde{y}}}) \cdot \frac{180}{2^{B_y}} - 90
\right] &\times\\
\left[
D + n_{\hat{z}},
D + \left(n_{\hat{z}} + 2^{B_z - l_{\tilde{z}}}
\right) \right]&
\end{split}
\end{align}

The numbers $n_{\hat{x}}, n_{\hat{y}}, n_{\hat{z}}$ can also be computed by right padding the respective bit string to $B_x, B_y$, and $B_z$ bits with zeros and then interpreting the value as an integer.
Analogously, the upper bounds for the volume can be computed by right padding them with ones, interpreting the result as an integer and increasing the value by one.

\end{appendices}

\end{document}